\documentclass[12pt]{article}%
\usepackage{amssymb}
\usepackage{amsmath}
\usepackage{amsthm}
\usepackage{amsfonts}
\usepackage{graphicx}
\usepackage{subfigure}%
\input epsf.tex
\begin{document}
\bigskip\begin{titlepage}
\begin{flushright}
UUITP-23/06\\
hep-th/0612222
\end{flushright}
\vspace{1cm}
\begin{center}
{\Large\bf The world next door\\}
\vspace{0.7cm}
{\Large Results in landscape topography\\}
\end{center}
\vspace{1cm}
\begin{center}
{\large
Ulf   H.\   Danielsson{$^1$}, Niklas Johansson{$^2$} and Magdalena Larfors{$^3$}} \\
\vspace{5mm}
Institutionen f\"or Teoretisk Fysik, Box 803, SE-751 08
Uppsala, Sweden \\
\vspace{5mm}
{\tt
{$^1$}ulf.danielsson@teorfys.uu.se\\
{$^2$}niklas.johansson@teorfys.uu.se\\
{$^3$}magdalena.larfors@teorfys.uu.se\\
}
\end{center}
\vspace{5mm}
\begin{center}
{\large \bf Abstract}
\end{center}
\noindent
Recently, it has become clear that neighboring multiple vacua might have interesting consequences for the physics of the early universe. In this paper we investigate the topography of the string landscape corresponding to
complex structure moduli of flux compactified type IIB string theory. We find that series of continuously connected vacua are common. 
The properties of these series are described, and we relate the existence of infinite series of
minima to certain unresolved mathematical problems in group theory. Numerical studies of the mirror quintic serve as illustrating examples.
\vfill
\begin{flushleft}
December 2006
\end{flushleft}
\end{titlepage}\newpage

\section{Introduction}

\bigskip

There are increasing evidence for the existence of a string landscape
describing a huge number -- possibly infinite -- of different vacua. Each of
these vacua corresponds to a low energy theory that, through string theory,
can consistently be completed into a full theory of quantum gravity. Possibly,
our universe is given by one of these vacua, and in that case we need to make
use of experiments and observations to find out which one. It is an
interesting question to ask in what way the presence of these other vacua
gives rise to directly or indirectly visible effects in our universe. Even if
the selection of our vacuum is to a large extent accidental, it could be
governed by anthropoid or statistical principles where the properties of the
landscape are probed. Through the use of conditional probabilities, and
carefully comparing probabilities for various combinations of the constants of
nature, one could find support for such a picture.

The presence of neighboring vacua could also have a more direct impact on the
physics of our universe. In order to be more specific, it is useful to use the framework
 of flux compactifications of type IIB
string theory. The moduli of the compact space -- typically given by Calabi--Yau
orientifolds -- give rise to scalar fields in four dimensional space time.
Fluxes on the compact space generate an effective potential for the moduli,
which then tend to be stabilized at the minima of the potential. Different
fluxes, and different minima for the same flux, correspond to different vacua
of string theory.

In \cite{Danielsson:2006jg} and \cite{Green:2006nv} it was argued that the
fixing of the moduli through fluxes can be affected through the presence of,
e.g., a black hole. An appropriately charged black hole generates an effective
potential for the moduli of the compact space in a way very similar to the
fluxes. This potential will compete with the potential generated by the fluxes
and the fate of the moduli will be determined by their sum. Far from the black
hole its contribution will be subdominant, but for a small black hole the
effects near the horizon could be substantial. As argued in
\cite{Danielsson:2006jg} for the case of moduli fixed near a conifold point
on the moduli space, the moduli could be shifted and lead to different four
dimensional physics. In \cite{Green:2006nv} it was argued that the presence of
a small black hole generated through a Hawking evaporation process, could lead
to a catastrophic transition to another vacuum. For these two effects to occur in
flux compactifications, there must exist at least two nearby vacua.

The presence of several nearby vacua could also have important consequences in
the early universe. In \cite{Tye:2006tg} it was argued that quantum resonance
effects in connection with tunneling could enhance the influence from one
vacuum on another. In \cite{Davoudiasl:2006ax} it was argued that domain walls
between regions of the early universe trapped in different vacua may lead to
effects visible even today. The domain walls collapse to black holes and could
generate a spectrum of primordial black holes surviving to this day. While a
speculative idea, it nevertheless gives an example of how the existence of
other vacua could lead to observable effects in our present day universe.

Another interesting proposal is the idea of chain inflation,
\cite{Freese:2006fk}. Instead of an inflaton rolling in an isolated minimum of
its potential, we have a series of minima with an inflaton tunneling from one
minimum to the next, reaching ever lower energies. In each minimum there will
be time for only a few e-folds but eventually the number of e-folds adds up to
the required value. 

Furthermore, a very interesting question has to to with the computability aspect of the selection problem. As emphasized in \cite{Denef:2006ad}, the string landscape bears a close resemblance with other landscapes like the landscape of all proteins or all living organisms. Here concepts such as the roughness of the landscape play important roles, as reviewed 
in \cite{Bryng:1995}.  Also, as shown in \cite{Acharya:2006zw}, sequences of connected vacua play an important role when studying the finiteness of the landscape. 

In all of the above cases it is important to map out the topographic properties of
the string landscape. How many minima do we have? How are they distributed?
What kind of barriers do we have? 

In this paper we focus on parts of the landscape where there are series of
continuously connected vacua.  Other examples of such considerations are given 
in \cite{Ceresole:2006iq} and \cite{Giryavets:2005nf}. We investigate the properties of these series
and the conditions for their existence. After a short review of relevant
features of flux compactifications in section two, we describe, in section
three, the main setup for our analysis. In section four we discuss various
types of series using explicit examples from the mirror quintic. In section
five we discuss the properties of the series we have discovered, and make a
connection with some so far unresolved mathematical problems. Finally, in
section six, we end with some conclusions and outlook.

\bigskip

\section{Background}
\label{CY_geom}

\bigskip

Flux compactifications of type IIB string theory on orientifolded Calabi--Yau
manifolds include models in which all moduli are dynamically stabilized.
Wrapped 3-form fluxes induce potentials for the complex structure moduli and
the axio-dilaton, \cite{DeWolfe:2002nn} and non-perturbative effects stabilize
the K\"{a}hler moduli \cite{Kachru:2003aw}. In previous works, such as
\cite{Kachru:2003aw}, the focus has been on the K\"{a}hler sector and the
behaviour of the potential for fixed complex structure. We concentrate instead
on the dependence on the complex structure moduli, which is better understood.
This provides a powerful framework for studying parts of the landscape of
string theory vacua. Below follows a brief review of relevant concepts in
Calabi--Yau geometry and flux compactifications\footnote{The geometrical
content of this section is covered by the seminal works \cite{Candelas:1990rm}
and \cite{Strominger:1990pd}. For a recent review on flux compactifications
with extensive references, see \cite{Grana:2005jc}.}.

\subsection{Calabi--Yau geometry}

Let us begin with some geometry. Let $X$ be a Calabi--Yau manifold with
complex structure moduli space $\mathcal{M}$. A key concept in the study of
Calabi--Yau moduli spaces are the period integrals - the \textquotedblleft
holomorphic volumes\textquotedblright\ of a basis of 3-cycles:
\begin{equation}
\Pi_{I}=\oint_{C_{I}}\Omega=\oint_{X}C_{I}\wedge\Omega.
\end{equation}
Here $\Omega$ is the holomorphic 3-form and $C_{I}$ denotes a basis of
$H_{3}(X)$ as well as its Poincar\'{e} dual. The index $I$ runs from $1$ to
$2h^{1,2}(X)+2\equiv N$. It is always possible to choose the basis $C_{I}$ so
that the intersection matrix $Q$ has the standard form
\begin{equation}
Q_{IJ}=\oint_{C_{I}}C_{J}=\oint_{X}C_{I}\wedge C_{J}=\left(
\begin{array}
[c]{ccc}%
0 & 0 & -i\sigma_{y}\\
0 & \cdots & 0\\
-i\sigma_{y} & 0 & 0
\end{array}
\right)  .
\end{equation}
The intersection matrix is left invariant under symplectic transformations. It
is customary and convenient to collect the periods into a vector
\begin{equation}
\Pi(z)=\left(
\begin{array}
[c]{c}%
\Pi_{1}(z)\\
\Pi_{2}(z)\\
\vdots\\
\Pi_{N}(z)
\end{array}
\right)  ,
\end{equation}
where $z$ is a $N/2-1$ dimensional (complex) coordinate on $\mathcal{M}$.

The space $\mathcal{M}$ is a topologically complicated complex space. This
manifests itself for instance in the fact that the periods (or, equivalently,
the 3-cycles) are subject to monodromies. Going around non-trivial loops in
$\mathcal{M}$ changes the periods by an integer symplectic matrix $T$:
\begin{equation}
\Pi\rightarrow T \cdot\Pi.
\end{equation}
All possible monodromy matrices constitute a group $M$ that is a subgroup of
$\mbox{Sp}(N,\mathbb{Z})$.

In our explicit example the mirror quintic we follow the conventions of
\cite{Denef:2001xn} (which are closely related to those of
\cite{Greene:2000ci}). Thus we parametrize $\mathcal{M}$ by a coordinate $z$
that takes values in the complex plane. In this plane there are two cuts: one
emanating from the large complex structure point $z = 0$ and one from the
conifold point $z = 1$. The monodromy matrices $T[0]$ and $T[1]$ around these
points generate the group $M$ and are given by
\begin{eqnarray}
T\left[  0\right]  =\left(
\begin{array}
[c]{cccc}%
1 & 1 & 3 & -5\\
0 & 1 & -5 & -8\\
0 & 0 & 1 & 1\\
0 & 0 & 0 & 1
\end{array}
\right) ,  &&  T\left[  1\right]  =\left(
\begin{array}
[c]{cccc}%
1 & 0 & 0 & 0\\
0 & 1 & 0 & 0\\
0 & 0 & 1 & 0\\
1 & 0 & 0 & 1
\end{array}
\right) , \label{eq:mono_mat}%
\end{eqnarray}
in the $\Pi$-basis of \cite{Denef:2001xn}. Thus, in these conventions, $C_{1}$
is the conifold cycle and $C_{4}$ is the cycle that intersects it. We use the
Meijer functions of the appendix of \cite{Denef:2001xn} to numerically
evaluate the periods $\Pi_{1}, \dots, \Pi_{4}$, except close to the conifold
where we make an expansion. Close to the large complex structure point $z=0$
we use the expansions of \cite{Denef:2001xn}.

\subsection{Flux compactifications}

Let us now turn to flux compactifications in which the internal manifold is an
orientifold of $X$. We use the same notations for the type IIB fields as
\cite{Giddings:2001yu}. In \cite{DeWolfe:2002nn} it was proved that wrapping
fluxes around the different 3-cycles of an orientifold of $X$ results in a
Gukov--Vafa--Witten superpotential:
\begin{equation}
W=\int_{X}\Omega\wedge(F_{(3)}-\tau H_{(3)})=F\cdot\Pi-\tau H\cdot\Pi\equiv
A+B\tau.\label{eq:W}%
\end{equation}
Here we collected the flux quanta of the RR flux into a vector $F$ defined
from $F_{(3)}=-F_{I}C_{I}=-F\cdot C$. Similarly for the NSNS fluxes:
$H_{(3)}=-H_{I}C_{I}=-H\cdot C$. We have rescaled $F_{(3)}$ and $H_{(3)}$ by a
factor $1/((2\pi)^{2}\alpha^{\prime})$, so that $F_{I}$ and $H_{I}$ are
integers. $W$ is accompanied by the K\"{a}hler potential
\begin{equation}
K=-\ln\left(  -i(\tau-\bar{\tau})\right)  +K_{\mathrm{cs}}\left(  z,\bar
{z}\right)  -3\ln\left(  -i(\rho-\bar{\rho})\right)  ,
\end{equation}
where $K_{\mathrm{cs}}$ is the K\"{a}hler potential for the complex structure moduli. The scalar potential for the complex structure moduli is given by the usual
$\mathcal{N}=1$ formula%
\begin{equation}
V\left(  z,\tau\right)  =e^{K}\left(  g^{i\bar{\imath}}D_{i}WD_{\bar{\imath}%
}\bar{W}+g^{\tau\bar{\tau}}D_{\tau}WD_{\bar{\tau}}\bar{W}+g^{\rho\bar{\rho}%
}D_{\rho}WD_{\bar{\rho}}\bar{W}-3|W|^{2}\right)  ,\label{superpot}%
\end{equation}
where $i$ goes over all complex structure moduli. Given that the K\"{a}hler
moduli can be stabilized simultaneously, minima of this potential correspond
to string theory vacua, and we may therefore use it as a tool for exploring
the landscape. We will consider both supersymmetric minima (for which
$D_{i}W=D_{\tau}W=0$) and minima lifted through F-terms, i.e. with non-zero
$D_{i}W$ and $D_{\tau}W$.

Perturbatively, the two last terms in (\ref{superpot}) cancel, producing a
no-scale potential. Depending on the details of the K\"{a}hler moduli
stabilization however, these terms might contribute through non-perturbative
effects. For instance, studying solutions with $D_{\rho}W = 0$ the potential
for the complex structure moduli and the axio-dilaton becomes
\begin{equation}
\label{superpot_nonoscale}V\left(  z,\tau\right)  = e^{K}\left(
g^{i\bar{\imath}}D_{i}WD_{\bar{\imath}}\bar{W}+g^{\tau\bar{\tau}}D_{\tau
}WD_{\bar{\tau}}\bar{W}- 3|W|^{2}\right)  .
\end{equation}
For simplicity, we focus on the no-scale case, but most of our results can be
generalized to a potential of the form (\ref{superpot_nonoscale}).

There is a tadpole cancelation condition on the fluxes that one can wrap on an
orientifold of $X$. Putting in $N_{D3}$ space-filling D3-branes, and having
$N_{O3}$ orientifold 3-planes, the condition reads
\begin{equation}
N_{D3} - \frac{1}{4}N_{O3} + \int_{X} H_{(3)} \wedge F_{(3)} = 0.
\label{eq:tadpole}%
\end{equation}
If the compactification is viewed as an F-theory compactification on a
four-fold $X_{4}$, the number of O3 planes is related to the Euler number of
$X_{4}$:
\begin{equation}
N_{D3} + \int_{X} H_{(3)} \wedge F_{(3)} = \frac{\chi(X_{4})}{24}.
\end{equation}
In terms of the matrix $Q$, the intersection product can be written
\begin{equation}
\int_{X} H_{(3)} \wedge F_{(3)} = H \cdot Q \cdot F.
\end{equation}
We now turn to a more detailed investigation of the scalar potential.

\bigskip

\section{The potential}

\bigskip

We set out to find minima of a potential of the form
\begin{equation}
V\left(  z,\tau\right)  =e^{K}\left(  g^{i\bar{\imath}}D_{i}WD_{\bar{\imath}%
}\bar{W}+g^{\tau\bar{\tau}}D_{\tau}WD_{\bar{\tau}}\bar{W}\right)
\label{scalarpot}%
\end{equation}
where $W$ is given by equation (\ref{eq:W}). If the K\"{a}hler moduli are
stabilized, such a minimum might correspond to our universe. In order to find
a minimum of the potential we need to find complex structure moduli $z$, and
an a dilaton $\tau$ such that $\partial_{i}V=\partial_{\tau}V=0$. The latter
condition can be put on a rather simple form:
\begin{equation}
\alpha(F,z)+\beta(F,H,z)\bar{\tau}+\gamma(H,z)\bar{\tau}^{2}=0,
\end{equation}
where $\alpha$, $\beta$ and $\gamma$ are all real functions of the fluxes and
the complex structure moduli given by%
\begin{align}
\alpha(F,z) &  =\left\vert A\right\vert ^{2}+g^{i\bar{\imath}}D_{i}%
AD_{\bar{\imath}}\bar{A}\\
\beta(F,H,z) &  =\bar{A}B+A\bar{B}+g^{i\bar{\imath}}D_{i}AD_{\bar{\imath}}%
\bar{B}+g^{i\bar{\imath}}D_{\bar{\imath}}\bar{A}D_{i}B\\
\gamma(H,z) &  =\left\vert B\right\vert ^{2}+g^{i\bar{\imath}}D_{i}%
BD_{\bar{\imath}}\bar{B}.
\end{align}
This determines $\tau=\tau\left(  z\right)  $ to be%
\begin{equation}
\tau(z)=-\frac{\beta(F,H,z)}{2\gamma(H,z)}\pm\sqrt{\frac{\beta(F,H,z)^{2}%
}{4\gamma(H,z)^{2}}-\frac{\alpha(F,z)}{\gamma(H,z)}},\label{eq:tau}%
\end{equation}
where, since $\alpha,\beta,\gamma$ are all real and one can show that the
expression under the square-root is negative semidefinite, $\tau$ has an
imaginary part given by the square-root term. The imaginary part of $\tau$ is
nothing but the inverse of the string coupling $g_{s}$, and a real $\tau$
therefore implies an infinite string coupling. Furthermore, a negative string
coupling is unphysical, so only the plus sign in the above equation gives
physically interesting values of $\tau$. Checking the second derivatives of
$V$ in the $\tau$ plane then shows that $V$ is always minimized in this plane.

Plugging the expression for $\tau$ back into the potential yields a function
$V(z,\tau(z))$ that only depends on the complex structure, and has an extremum
exactly when $\partial_{i} V = 0$:
\begin{equation}
\frac{dV\left(  z,\tau\left(  z^{k}\right)  \right)  }{dz^{i}}=\frac{\partial
V\left(  z,\tau\left(  z^{k}\right)  \right)  }{\partial z^{i}}+\frac
{d\tau\left(  z^{k}\right)  }{dz^{i}}\frac{\partial V\left(  z,\tau\left(
z^{k}\right)  \right)  }{\partial\tau}= \frac{\partial V\left(  z,\tau\left(
z^{k}\right)  \right)  }{\partial z^{i}}.
\end{equation}
In the plots of the potential for the mirror quintic, it is the function
$V(z,\tau(z))$ that is plotted unless explicitly mentioned otherwise.

Let us now make a few remarks concerning the structure of the scalar potential
$V$. Being defined in terms of the periods, $V$ is not singled valued on the
complex structure moduli space, but only on its cover, the Teichm\"{u}ller space.
Specifically, going around a non-trivial loop in moduli space will in general
change the potential, $V\rightarrow\tilde{V}$. Physically, this change is due
to monodromies of the cycles that the fluxes wrap: going around the loop, we
return to a manifold that looks the same, but on which the fluxes wrap
different cycles. The potential changes continuously, but does not return to
its original value.

As mentioned in section \ref{CY_geom}, the monodromies of the cycles (or
equivalently of the periods) are formulated in terms of matrices. Going around
a non-trivial loop in moduli space, e.g. around a conifold point, the periods
change according to
\begin{equation}
\Pi\rightarrow T \cdot\Pi,
\end{equation}
where $T$ is an $N$ by $N$ integer symplectic matrix. $V$ changes since the
superpotential does: $W = (F - \tau H) \cdot\Pi\rightarrow(F-\tau H) \cdot T
\cdot\Pi$. We find it more convenient to keep the periods fixed and instead
transform the flux vectors: $F \rightarrow F \cdot T$ and $H \rightarrow H
\cdot T$.

\begin{figure}[tb]
\centering
\includegraphics[height=12cm]{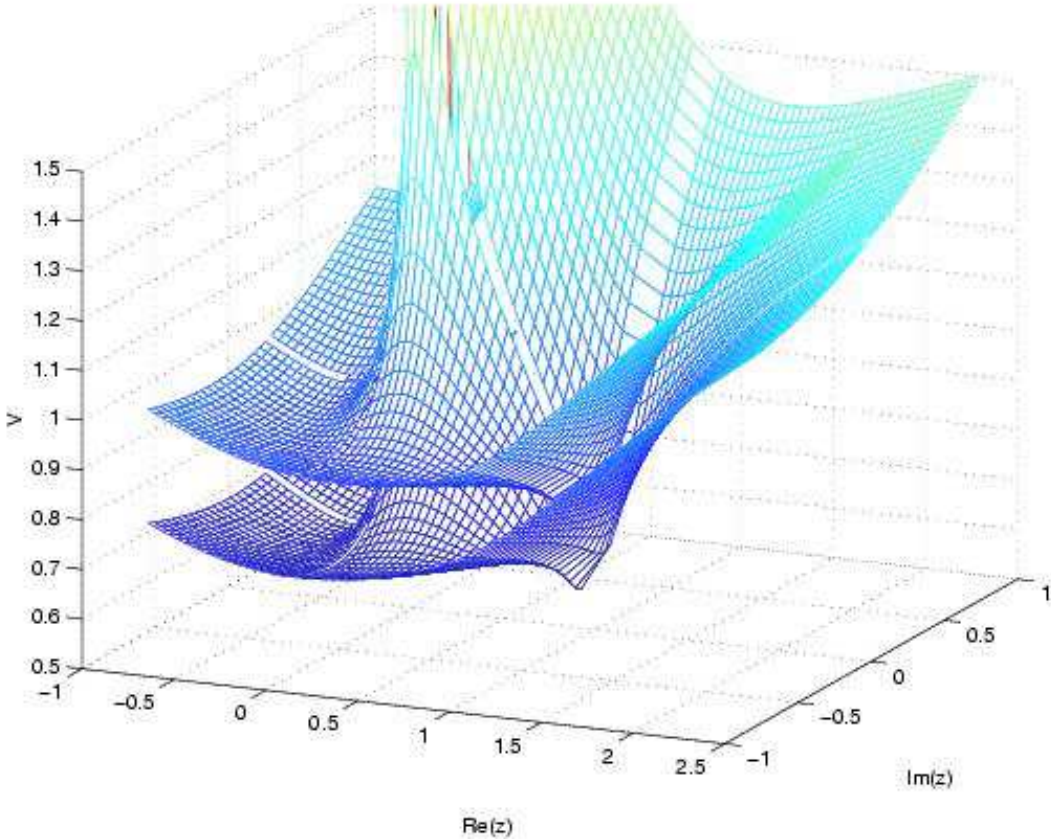} \caption{{\small \textsl{The scalar
potential $V$ is a multivalued function on moduli space. In the figure two
sheets of this function, each containing a minimum, are plotted. Note that
there is a smooth path connecting the two minima.}}}%
\label{Riemann}%
\end{figure}

As we will argue, moving around a fix-point of a monodromy allows for
continuously connected sequences of physically distinguished minima of the
scalar potential. Considering a fundamental domain in the
Teichm\"{u}ller space having a cut extending from the fix-point, the picture is as
follows. Let us assume a specific flux configuration such that a minimum is
localized in the fundamental domain. The minimum need not, necessarily, be
close to a fix-point. If we then move away from the minimum through the cut
extending from the fix-point we transform the fluxes accordingly to find out
how the potential continues after the cut. It turns out that it is often the
case that one again finds a minimum. Note that this is a new minimum,
physically different from the first, and that there is a continuous way of
going from one minimum to another with an intervening potential barrier.
Figure \ref{Riemann} gives an explicit example of this. 

\bigskip

\section{Series of connected minima}

\bigskip

We will now be a bit more specific, and consider minima related by monodromies
around the different fix-points in the complex structure moduli space. Our
discussion will be completely general, but we will use the mirror quintic to
find illustrating examples of our results. We start with the most transparent
case: the monodromy around a conifold point, which in the mirror quintic case
has the monodromy matrix (\ref{eq:mono_mat}).

Assume a superpotential given by (\ref{eq:W}), and apply $n$ consecutive
monodromies around a conifold point. If we choose the shrinking cycle to be
$C_{1}$, and the intersecting cycle as $C_{N}$, the fluxes are transformed
as:
\begin{align}
F  &  \rightarrow F+n\left(  F_{N},0,0,...,0\right) \\
H  &  \rightarrow H+n\left(  H_{N},0,0,...,0\right)  .
\end{align}
Here $F_{N},H_{N}$ are the
fluxes wrapping $C_{N}$. It is easy to see that the fluxes will be dominated
by the components $F_{N},H_{N}$ after sufficiently many monodromies.

We start by looking at an example of a series of minima for this general case.
Before we do so, we will simplify the problem, using the fact that type IIB
theory is invariant under SL(2,$\mathbb{Z}$) transformations of the axio-dilaton and the
fluxes \cite{Giddings:2001yu}. It is easy to show that the general flux configuration above can
always be transformed to a configuration where $H_{N}=0$. The transformed configuration is somewhat easier to analyze under monodromies, since only $F$ change.

\begin{figure}[p]
\centering
\subfigure{\includegraphics[height=8cm]{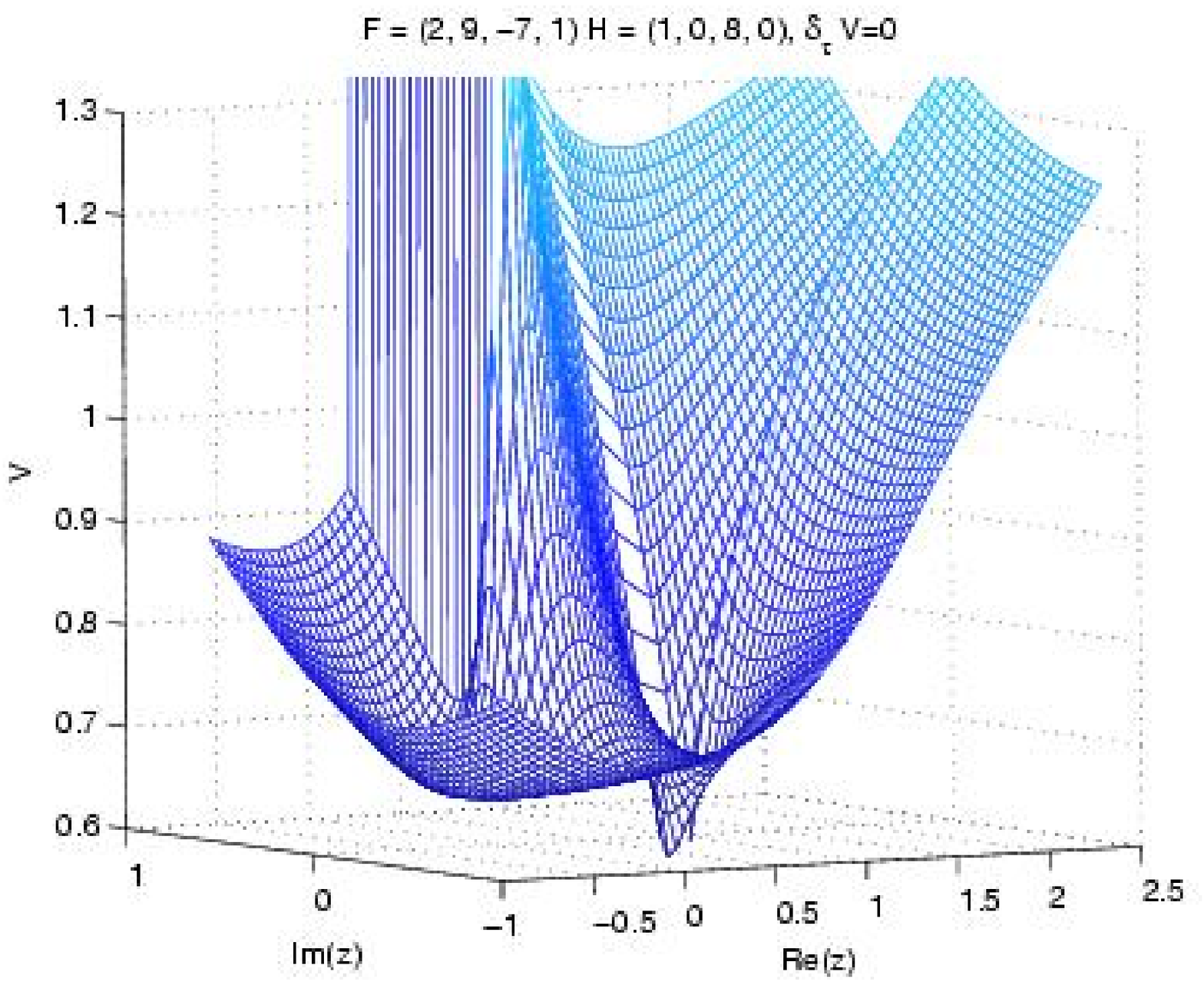}} \quad
\subfigure{\includegraphics[height=8cm]{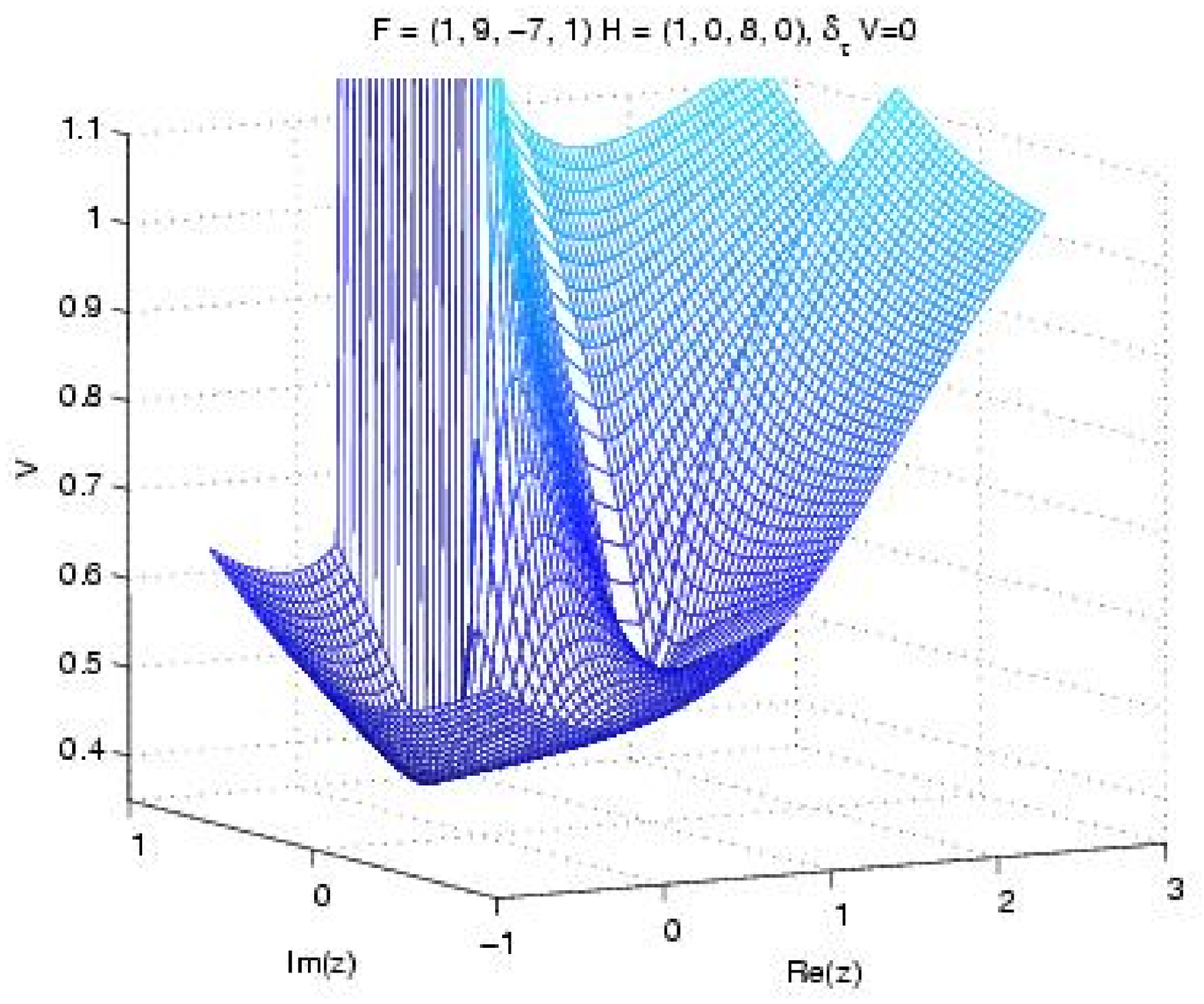}} \caption{{\small \textsl{Two
minima related by a conifold monodromy. Note that there is no minimum at the
conifold point $z=1$ in the figures, since there is a cut extending from this
point. The minima are located at $z=-0.31-0.17i$ and $z=-0.15+0.15i$
respectively. Going under the cut in the upper figure (from positive to
negative $\operatorname{Im}(z)$), we end up on the ``up-side'' of the cut on the lower
figure (negative $\operatorname{Im}(z)$).}}}%
\label{fig:serie1}%
\end{figure}

\subsection{A general example}

As an explicit example we will study the flux configuration $F =(2,9,7,-1)$
and $H=(1,0,8,0)$ on the mirror quintic. As shown in figure \ref{fig:serie1},
the corresponding potential has a minimum in the fundamental domain of the
Teichm\"{u}ller space. Now we perform an inverted conifold monodromy,
$T[1]^{-1}$, moving down through the cut extending from the conifold point.
Thus we end up in a new fundamental domain, or, equivalently, in the same
domain but with changed flux quanta: $F=(1,9,7,-1)$ and $H=(1,0,8,0)$. It is
to be expected that such a small change in the fluxes only changes the
potential by a small amount. Indeed, in figure \ref{fig:serie1} we see that
the potential looks more or less the same, but that the minimum has moved and
the value of the potential at the minimum is slightly lower.

\begin{figure}[p]
\centering
\subfigure{\includegraphics[height=8cm]{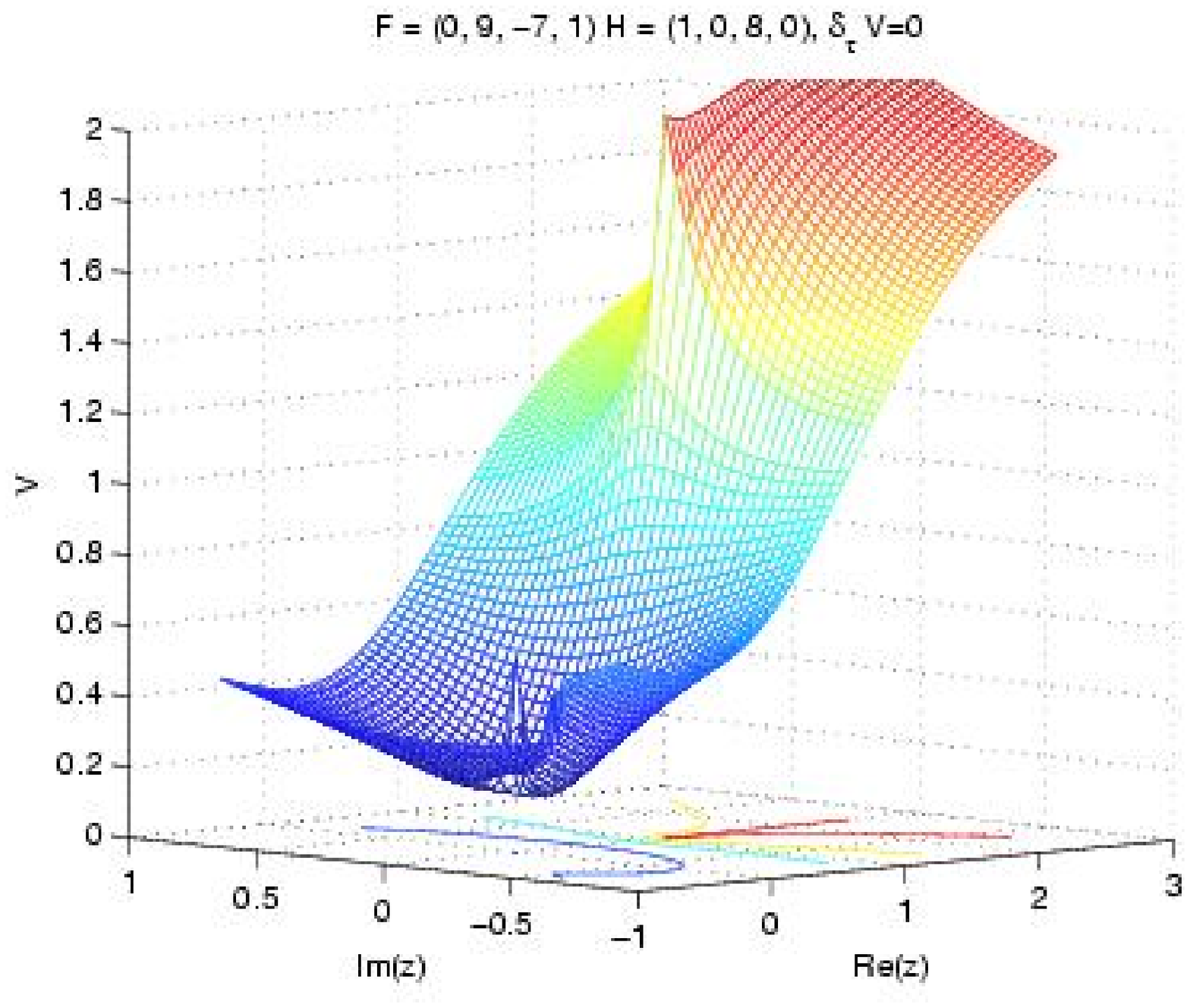}} \quad
\subfigure{\includegraphics[height=8cm]{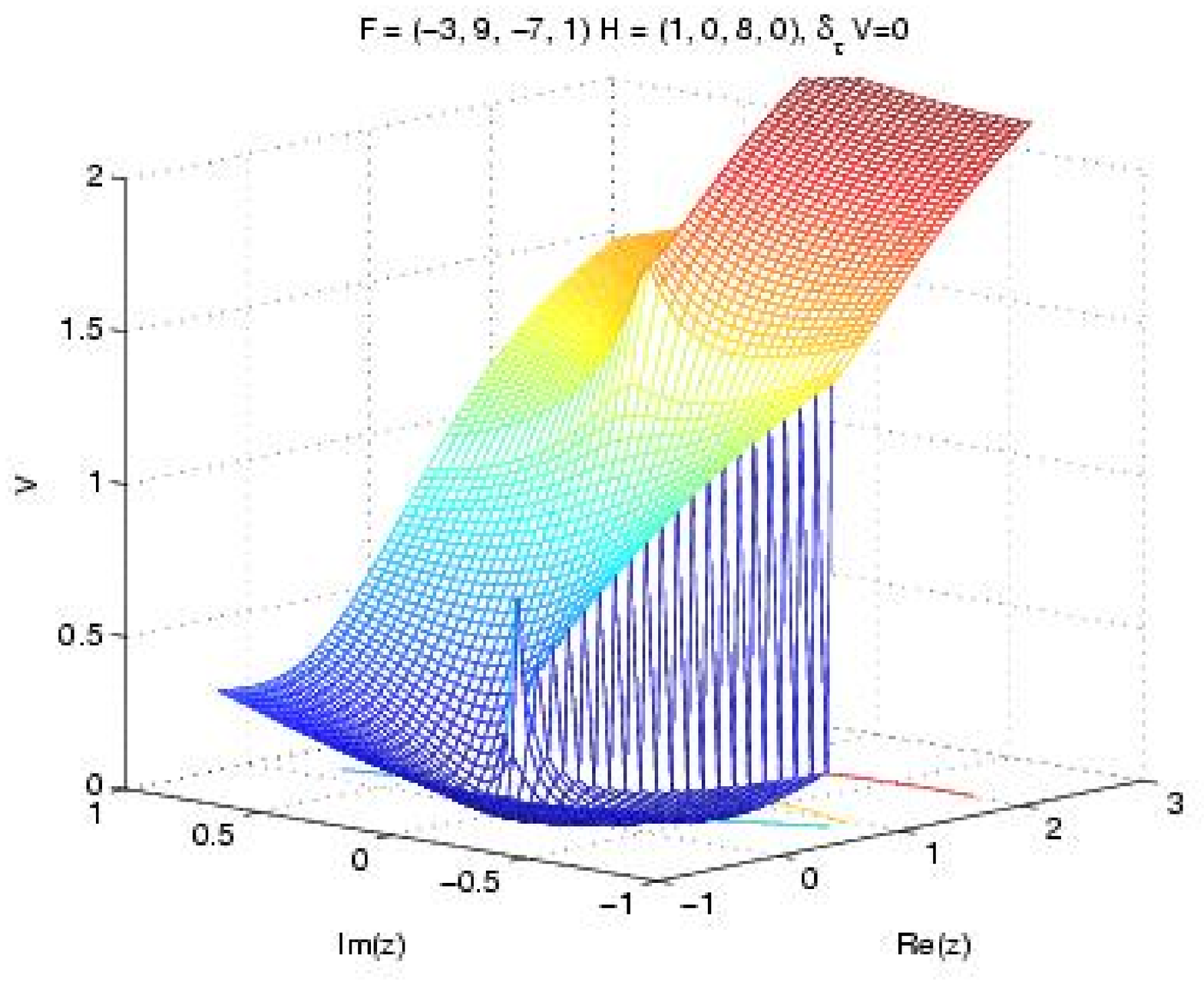}}
\caption{{\small \textsl{The third and the sixth minima of the series. Note
that the cut extending from $z=0$ has to be moved, in order to trace the
minima. Note that the sixth minimum is supersymmetric.}}}%
\label{fig:serie2}%
\end{figure}

As we continue applying monodromies, the minimum starts encircling the large
complex structure point ($z=0$ in the figures) clockwise. This means that the
minimum approaches the cut extending from this point, eventually crossing it.
In order to trace the minimum, we move this cut by performing a monodromy
around $z=0$. Having moved the cut, we find new minima with decreasing minimal
value, as shown in figure \ref{fig:serie2}.

We continue in this fashion. As $-F_{1}$ increases, the minimum moves further
and further away, yielding a different value of the complex structure modulus.
But apart from this, the potential is more or less similar to the second
picture in figure \ref{fig:serie2}. The extremal value of the potential,
however, goes down to zero, so the minima become supersymmetric\footnote{To
check whether the minmal value is exactly zero, we fixed $\tau$ so that
$D_{\tau} W = 0$ and then plotted lines where the real and imarinary parts of
$D_{z} W$ change sign. If these lines cross, the minimum value must be zero.}.
In this way, we can trace nine similar minima encircling the large complex
structure moduli point counterclockwise. After nine monodromies, we notice a
new feature. When $F_{1}=-13$ and $-14$, there are no visible minima in our
plots. Instead there is a spiral around the large complex structure point,
$z=0$. However $F_{1}=-15$ again produces a supersymmetric minimum.

\begin{figure}[tbh]
\centering
\subfigure{\includegraphics[height=5.6cm]{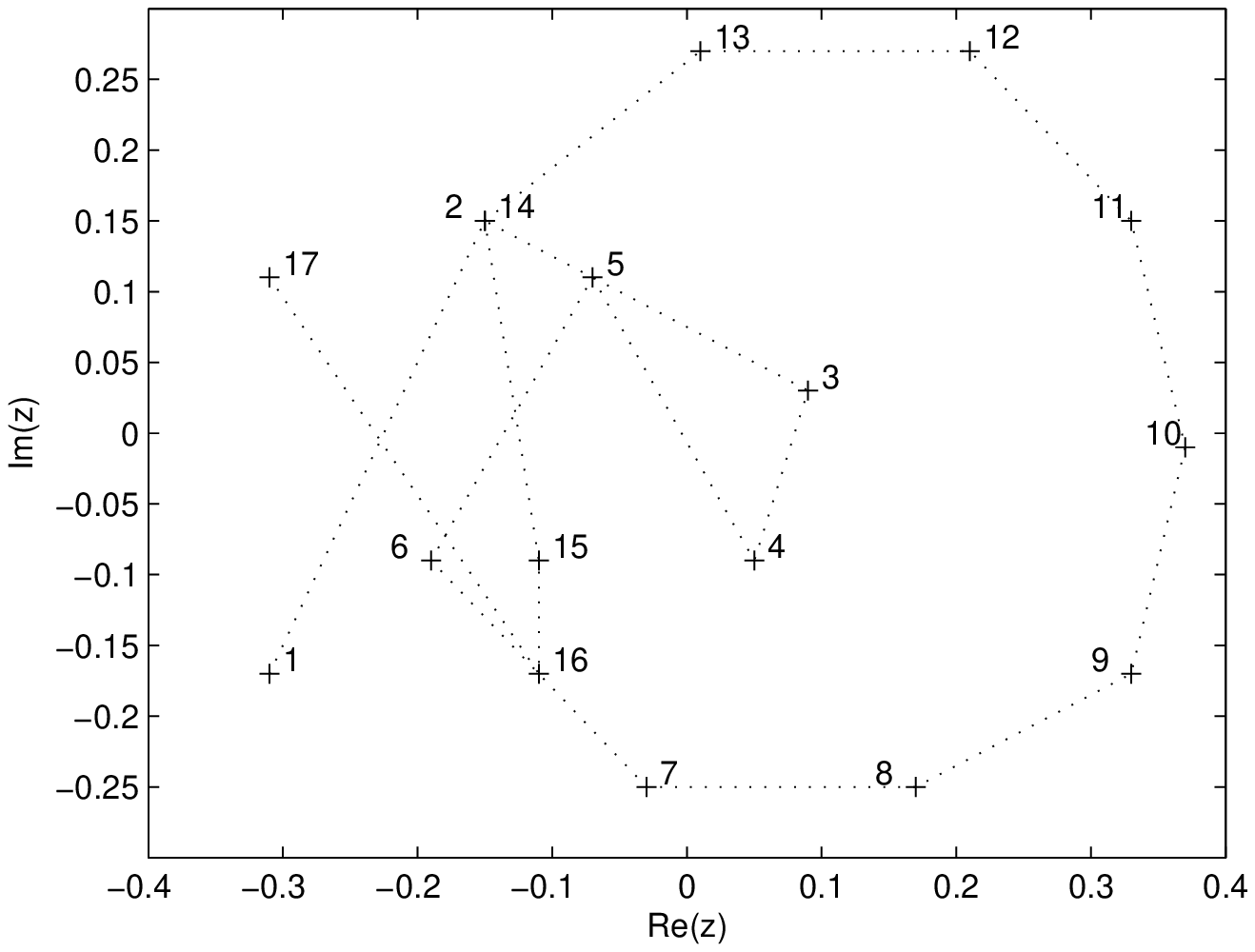}}
\subfigure{\includegraphics[height=5.6cm]{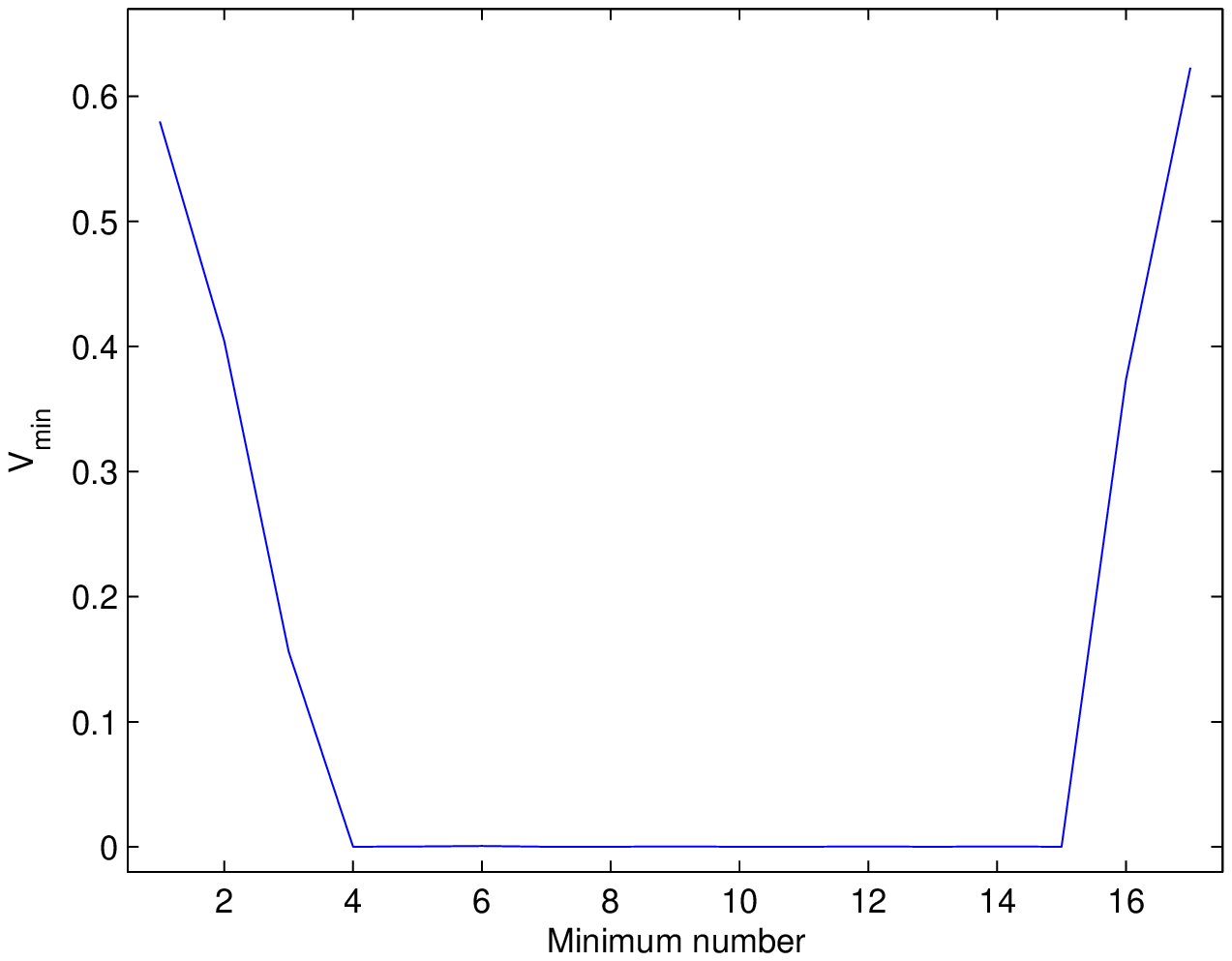}}
\caption{{\small \textsl{The position and extremal value of minima related by
$T[1]^{-1}$. The flux quanta of the first minimum are $F=(2,9,7,-1)$ and
$H=(1,0,8,0)$. The subsequent minima have decreasing values of $F_{1}$, as
described in the text. There are twelve connected vacua that are
supersymmetric.}}}%
\label{fig:serie3}%
\end{figure}

Proceeding with the monodromies, the potentials now change appearances. The
minima become more shallow, and the value of the potential at the minimum
increases. Eventually, the potential again looks like the first minimum in our
series. After that, the minimum disappears altogether, and the picture is
dominated by a funnel centered at the conifold. We will discuss the behaviour
around the conifold further down. The positions of the minima and the extremal
values of the potentials are shown in figure \ref{fig:serie3}. Note that
$F_{1}=-13$ and $-14$ are not included in the plots, since they do not
correspond to any minima.

\subsection{Periodic series}

Digressing from the general form of the fluxes, we note that another
intriguing feature appears when we consider fluxes of the form
\begin{align}
F  &  =\left(  F_{1},F_{2},F_{3},...,F_{N}\right) \\
H  &  =\left(  H_{1},0,0,...,0\right)  .
\end{align}
These transform as $F \rightarrow F+n\left(  F_{N},0,0,...,0\right)  , H
\rightarrow H$ under $n$ conifold monodromies. A simple calculation shows that
the axio-dilaton transforms into
\begin{equation}
\tau_{0}\rightarrow-n\frac{F_{N}}{H_{1}}+\tau_{0}. \label{eq:trans}%
\end{equation}
Here $\tau_{0}$ is the value $\tau$ has before the monodromies. Thus, the real
part of $\tau$, the axion, is shifted by a (not necessarily integer) factor
$n\frac{F_{N}}{H_{1}}$, whereas the dilaton is unaffected.

In the superpotential the shift in $\tau$ is canceled by the shift in the
fluxes. Similarly, the potential also stays the same. Hence, if the potential
has a minimum in the fundamental domain, it will have a second minimum after
the monodromy, that only differs from the first by the value of the axion and
the value of the $F$ flux. If the initial minimum fulfills the requirements of
a string theory vacuum, then so will the final minimum, even after an infinite
number of monodromy transformations. It seems that we have found an infinite
number of vacua, connected by continuous paths in the Teichm\"{u}ller space of the
complex structure moduli.

\begin{figure}[p]
\centering
\subfigure{\includegraphics[height=8cm]{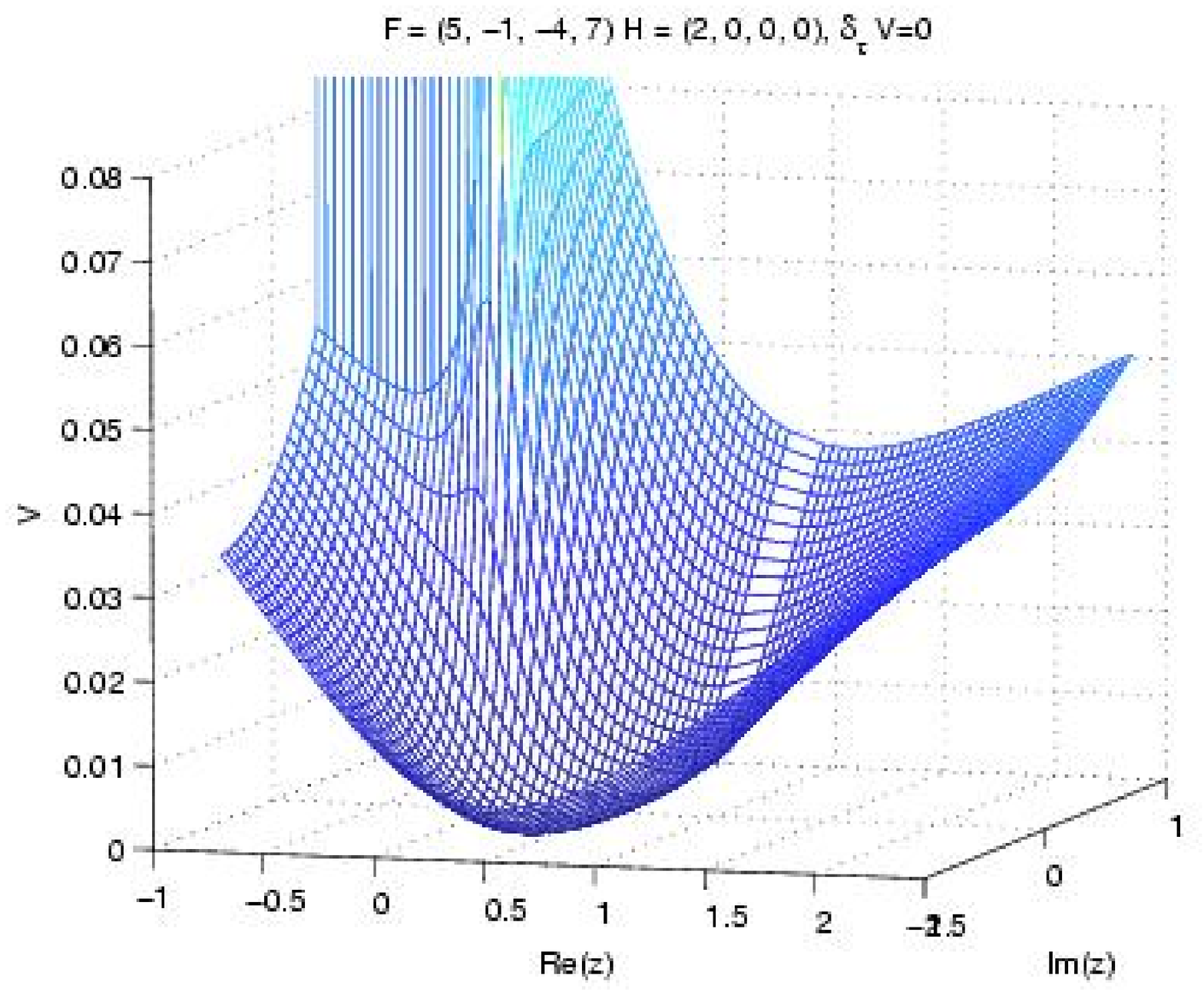}} \quad
\subfigure{\includegraphics[height=8cm]{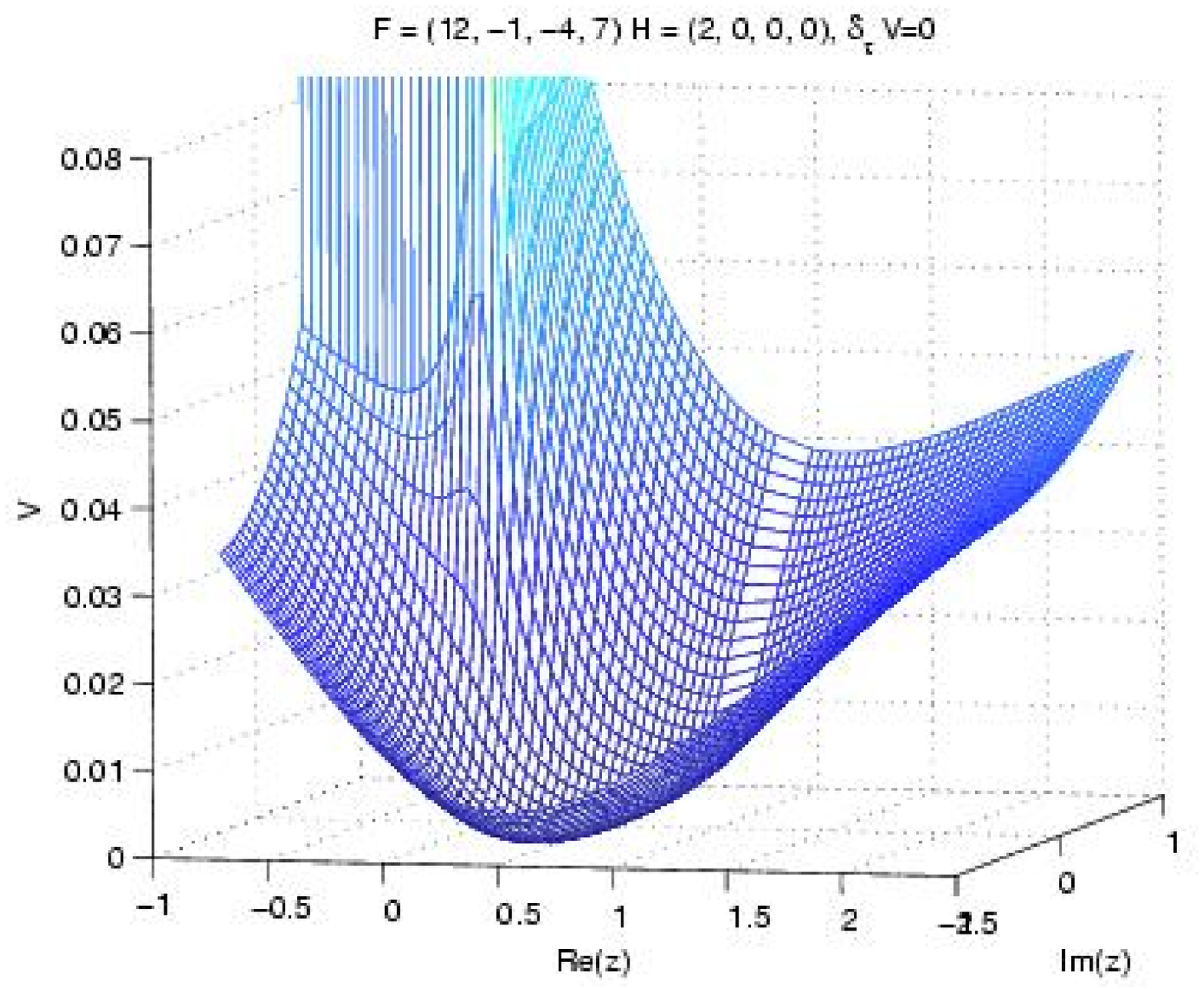}}
\caption{{\small \textsl{If the flux quanta are chosen such that
$H=(H_{1},0,0,0)$, then minima related by a conifold monodromy are identical
apart from the value of the axion. Note that the flux configurations differ
between the two plots.}}}%
\label{fig:axion1}%
\end{figure}

However, this series is periodic. $F_{N}$ and $H_{1}$ are integers because of
the Dirac quantization conditions. Thus, for some $n$, $n \frac{F_{N}}{H_{1}}$
will be integer and the monodromy combined with the transformation
(\ref{eq:trans}) will simply be the well-known SL(2,$\mathbb{Z}$) symmetry of
type IIB supergravity.\footnote{There are of course cases where $F_{N} = k
H_{1}$, for some integer $k$ already to begin with, but the point we want to
make is that other possibilities exist.} This shift symmetry is connected to
the periodicity of the axion \cite{Conlon:2006tq} and shows that the series of minima is indeed
finite. Even so, we note that, by choosing $F_{N}$ and $H_{1}$ to be
relatively prime, the period of the axion can be made rather large ($= H_{1}%
$), so it is possible to get an accretion of vacua at specific points in the
fundamental domain of the complex structure moduli space. Furthermore, such a
series of minima is very interesting from a topographic point of view, as
pointed out in the introduction.

\begin{figure}[p]
\centering
\subfigure{\includegraphics[height=8cm]{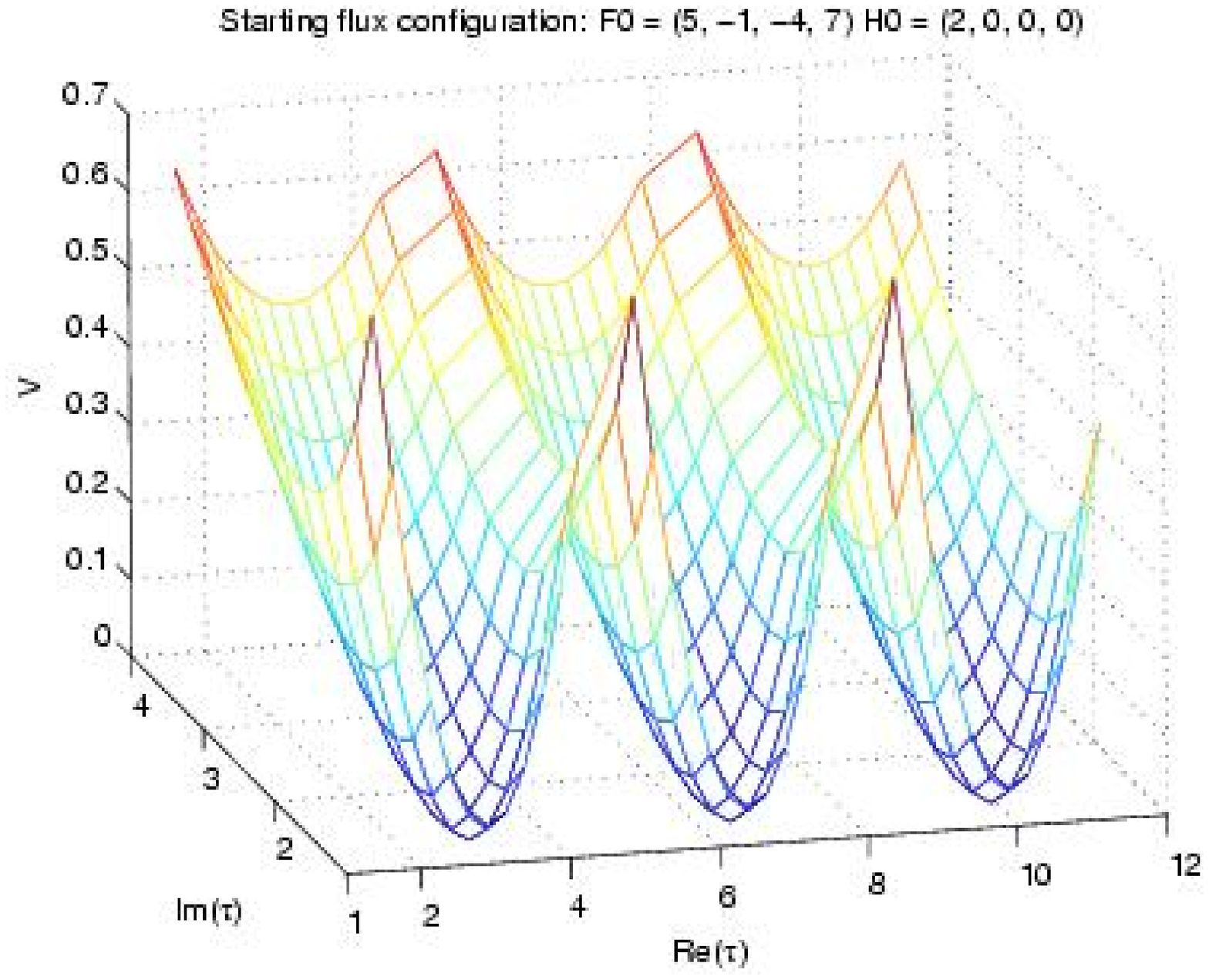}} \quad
\subfigure{\includegraphics[height=8cm]{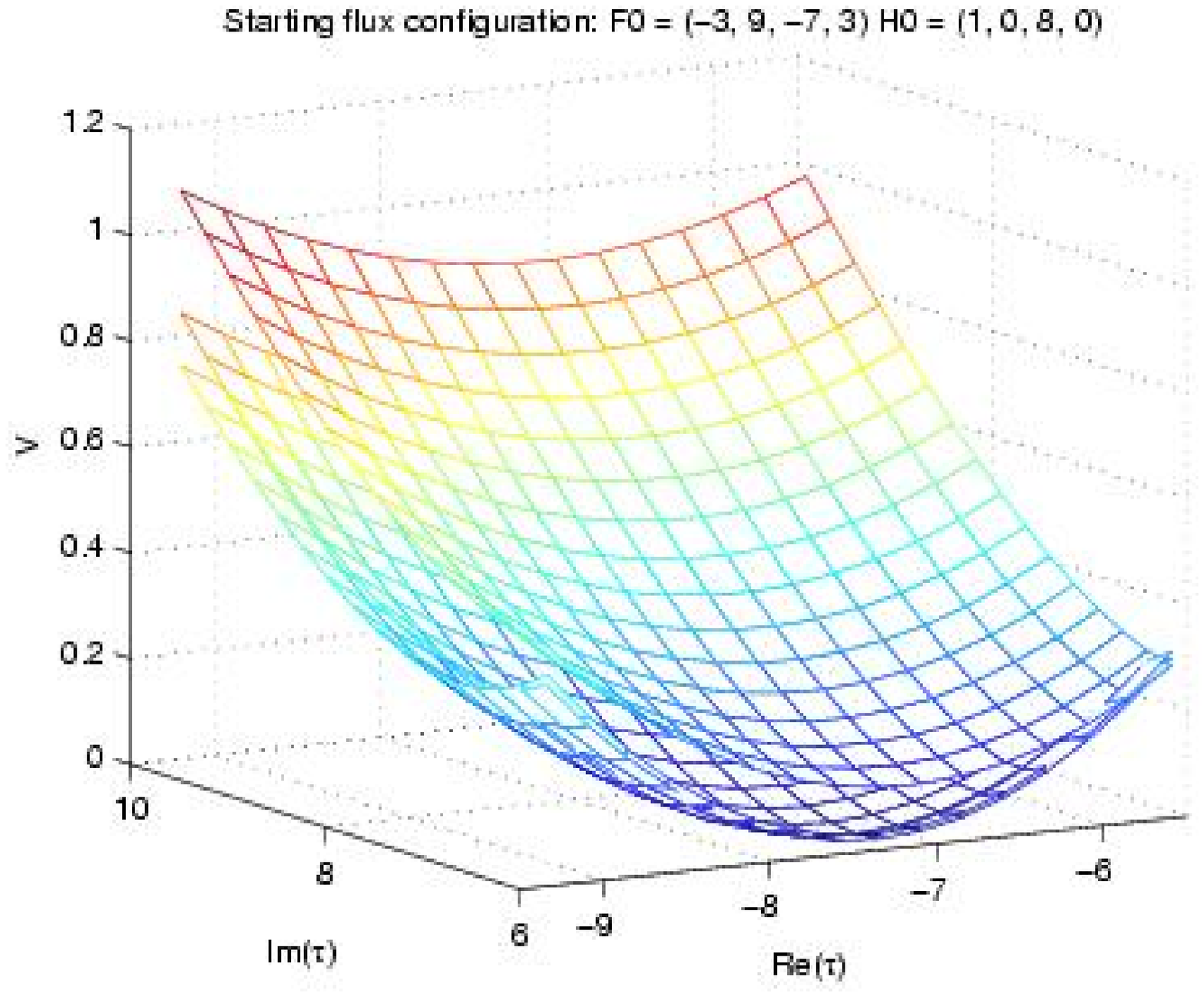}}
\caption{{\small \textsl{The potential plotted around the minima in the $\tau
$-plane. Note the difference between the periodic and the non-periodic case
when changing the flux quanta by conifold monodromies. Each sheet, or
potential well, corresponds to different fluxes.}}}%
\label{fig:axion2}%
\end{figure}

There is one more interesting special case where minima are connected in a
similar fashion. This is when the neither flux has a component through $C_{N}%
$, i.e. $F_{N}=H_{N}=0$. In these cases a monodromy leaves every parameter of
the vacuum unchanged, including the fluxes. We thereby get a series of
equivalent vacua connected by continuous paths. These vacua should be identified, 
since they all lie at the same place in the
combined space of fluxes and moduli. However, as discussed above, the
topography of the landscape is changed, i.e. the existence of paths connecting
equivalent minima yields a different situation than if we have one isolated
minimum, which makes these series of identical minima interesting in their own right.

\subsection{Limiting flux configurations}

Let us now study what the potential looks like after a large number of
monodromies. Is it possible to find infinite sequences of connected minima?
We return to completely general fluxes, but for simplicity we partially fix
the SL(2,$\mathbb{Z}$) symmetry by setting $H_{N}=0$, i.e.,
\begin{align}
F  &  =\left(  F_{1},F_{2},F_{3},...,F_{N}\right) \\
H  &  =\left(  H_{1},H_{2},...,H_{N-1},0\right) .
\end{align}
As $n$ goes to infinity one can show that $\operatorname{Im}(\tau) \sim n$ and $\operatorname{Re}(\tau)
\sim1/n^{2}$. Thus, we can find a series of potentials where as $n\rightarrow
\infty$, we have $g_{s}\rightarrow0$. This looks promising for finding long
series of vacua in the full string theory.

For the potential and the normalized superpotential $e^{K/2}W$ we get
$V\sim\mathcal{O}(n)$ and $e^{K/2}W\sim\mathcal{O}(\sqrt{n})$. In the
large-$n$ limit, the fluxes are
\begin{align}
F  &  \rightarrow F+n \left(  F_{N},0,0,...,0\right)  = F + n F_{L}\\
H  &  \rightarrow H = H_{L},
\end{align}
In order to generate an infinite series of vacua of the above form, we should
have limiting fluxes $F_{L}, H_{L}$ that yield a minimum for the
potential $V_{L}$ corresponding to $F_{L}$ and $H_{L}$. This follows from a
straight-forward calculation of $V$:
\begin{equation}
V = nV_{L} + \mathcal{O}(1),
\end{equation}
so if $V_L$ has a minimum, so will $V$, as $n \rightarrow \infty$.

One can show that, for a general Calabi--Yau, $F_{L},H_{L}$ will always
generate a minimum at the conifold point. Near the conifold where $z=1$ we
have, if we introduce the coordinate $\xi=z-1$,%
\begin{equation}
K_{\xi\bar{\xi}}=-\frac{1}{2\pi}\ln\xi\bar{\xi}+...
\end{equation}
This implies that%
\begin{equation}
\tau=nF_{N}\left(  -\frac{D_{\xi}\Pi_{1}\overline{D_{\xi}B}+\overline{D_{\xi
}\Pi_{1}}D_{\xi}B}{2\left\vert B\right\vert ^{2}K_{\xi\bar{\xi}}}+\frac
{i}{\left\vert B\right\vert }\frac{1}{\sqrt{K_{\xi\bar{\xi}}}}\right)
+\mathcal{O}(\xi)
\end{equation}
Even though there is an overall $n$ making $\tau$ large, there is a competing
effect making $\tau$ small if we move closer to the conifold point. We
furthermore find that the potential goes to zero like%
\begin{equation}
V\sim\frac{1}{\sqrt{K_{\xi\bar{\xi}}}}%
\end{equation}
at the conifold point. Thus, $V_{L}$ always has a minimum\footnote{Note that this is not 
an extremal point of the potential since $\partial_{\xi} V_{L}$ is singular.}. However, this
minimum can never be reached by monodromies as we now explain.

\begin{figure}[tb]
\centering
\includegraphics[height=12cm]{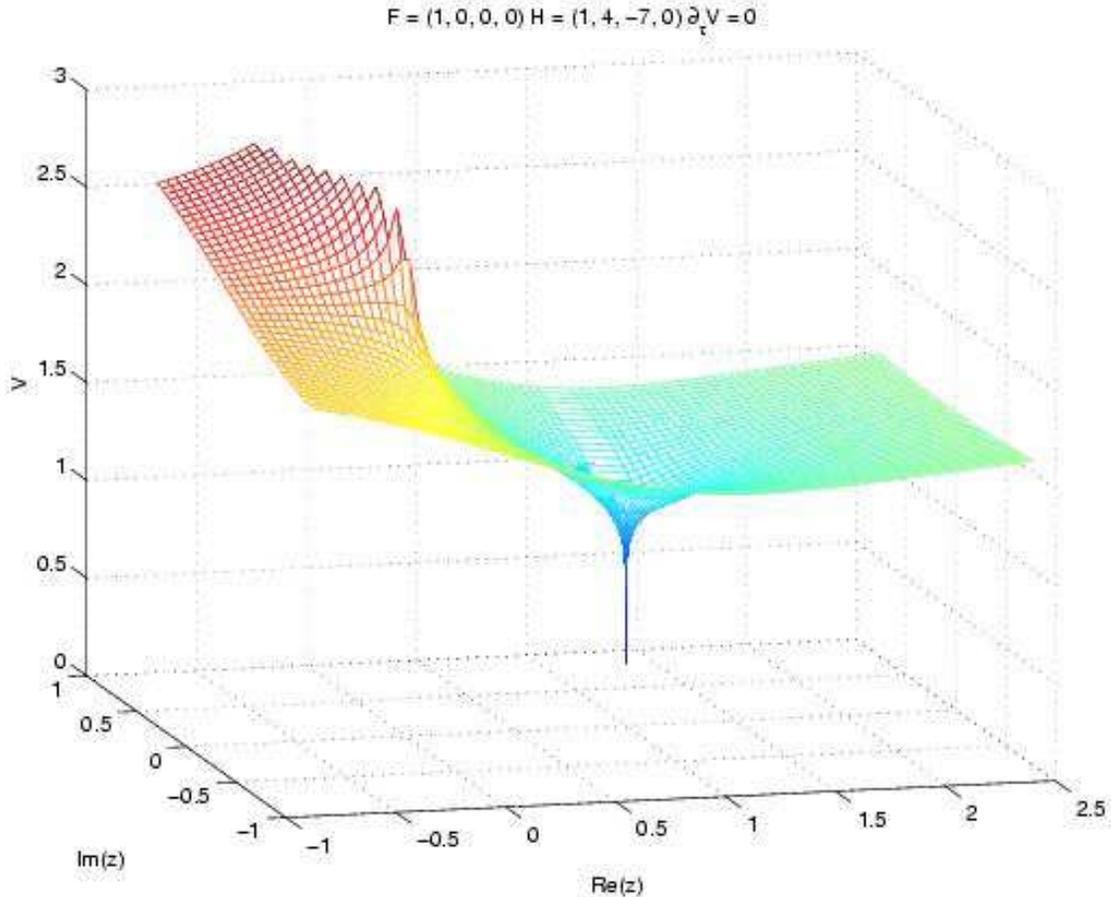} \caption{{\small \textsl{The
limiting flux configuration $F_{L},H_{L}$ always has a very deep minimum at
the conifold point $z=1$.}}}%
\label{fig:con_grans}%
\end{figure}

\begin{figure}[p]
\centering
\centering
\subfigure{\includegraphics[height=8cm]{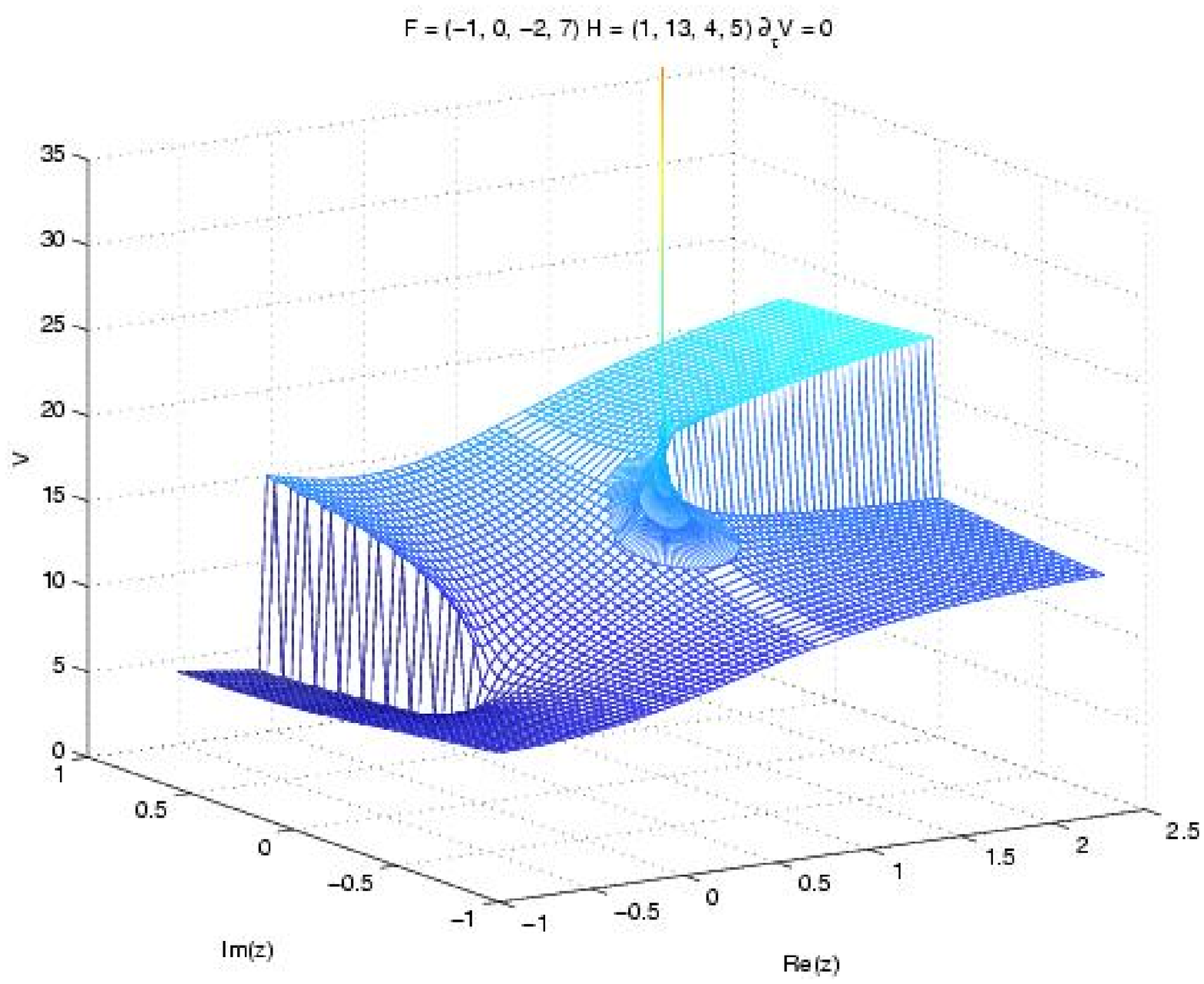}} \quad
\subfigure{\includegraphics[height=8cm]{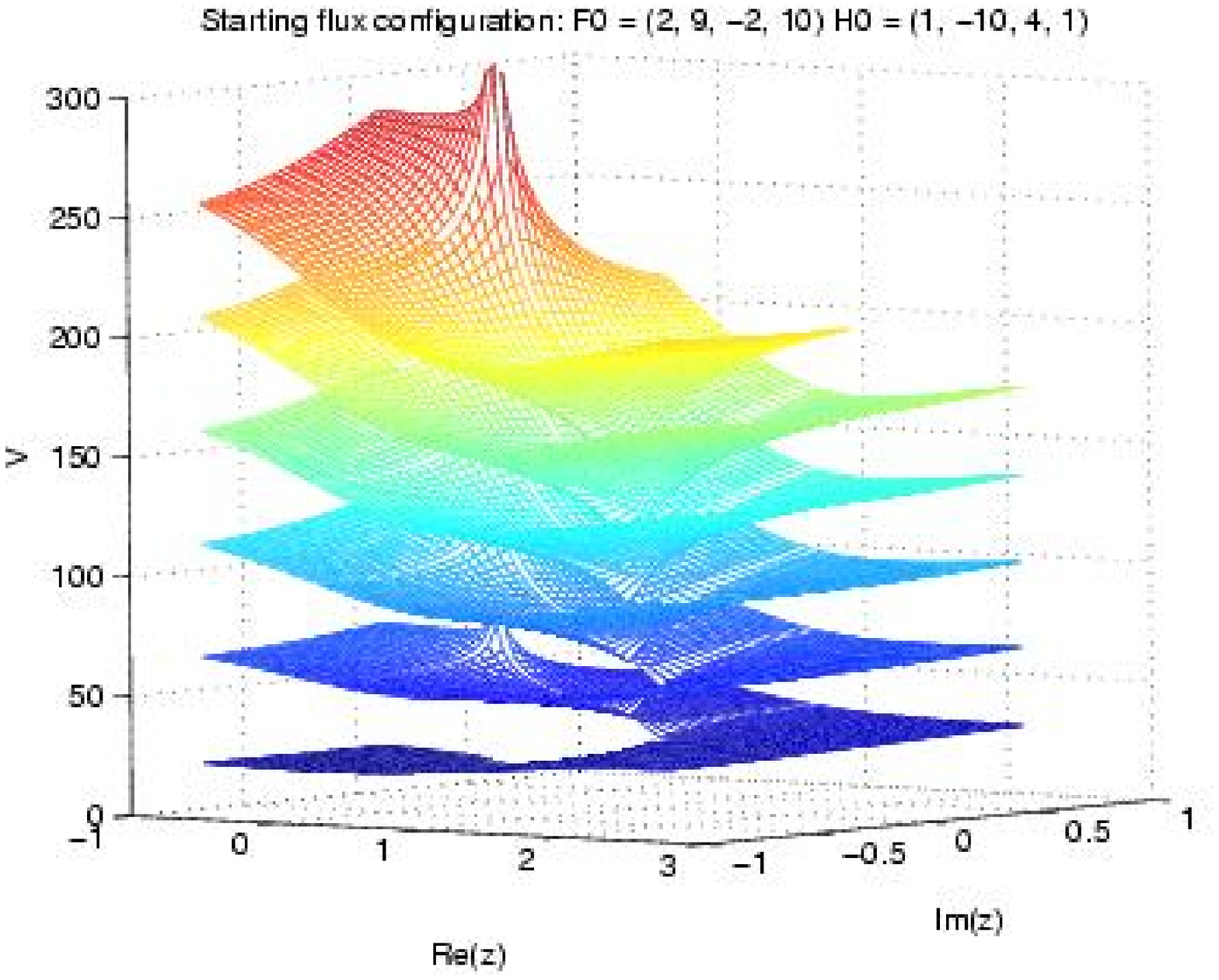}}
\caption{{\small \textsl{Approaching the limiting flux configuration, the
potential looks very different compared to figure \ref{fig:con_grans}. There
is a cut extending from the conifold and a spike centered at the conifold.
Plotting several sheets of the potential, we see that it looks like a spiral
staircase.}}}%
\label{fig:stair_spike}%
\end{figure}

In the limiting potential there are no cuts around the conifold point, see
figure \ref{fig:con_grans}. Thus the potential is unchanged by monodromies
around this point. On the way towards the limiting expression, however, there
is in general a nonzero $F_{N}$ and $H_{N}$ producing a cut with a decreasing
relative height, and an infinite spike in the middle, as shown in figure
\ref{fig:stair_spike}. In the above derivation with a vanishing potential at
the conifold point it was crucial that no such terms were present.

In order to reach the limiting potential we go through the cut, entering a
spiral stairway going either further upwards or downwards, as shown in figure
\ref{fig:con_grans}. In order to reach the funnel-looking minimum for the
limiting fluxes we climb upwards. However, we will always have a cut in the
potential, no matter how far we climb. Thus we can never reach the limiting
minimum at the top of the stairs by monodromies around the conifold point.

Nevertheless, if we find limiting fluxes that yield a minimum at some other
point in the complex structure moduli space, we would find an infinite series
of minima. As $n$ increases, the potentials would be more and more similar to
the limiting case. In particular, the local structure around the minimum would
be alike, apart from a scaling of the potential (recall that $V\sim n$
for large $n$). The puzzling thing is that we have not found such minima in
the limiting case for the mirror quintic. This suggests that infinite series
of minima connected by monodromies either do not exist or are very uncommon.
We will return to this question in section 5.

\begin{figure}[p]
\centering
\centering
\subfigure{\includegraphics[height=8cm]{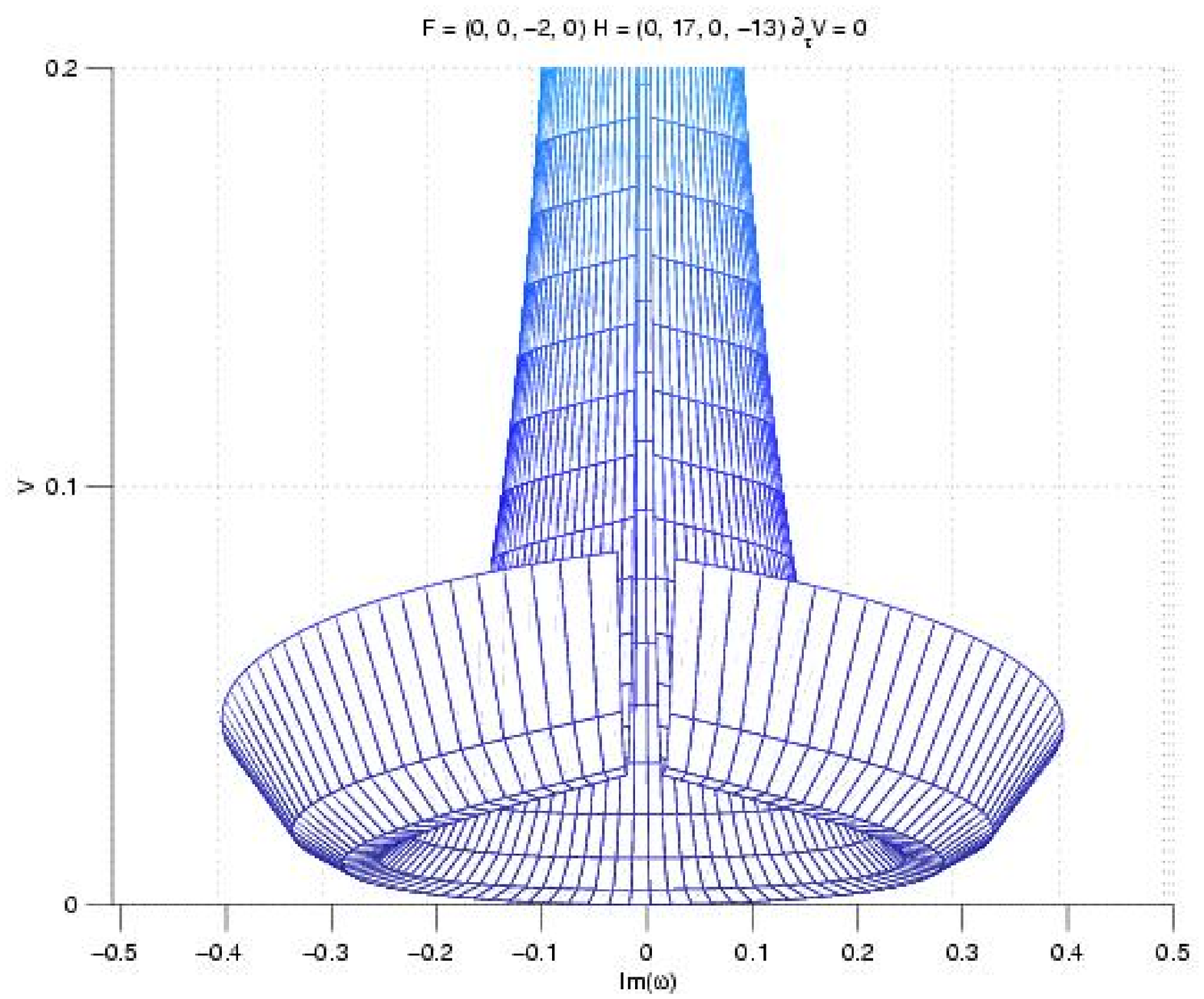}} \quad
\subfigure{\includegraphics[height=8cm]{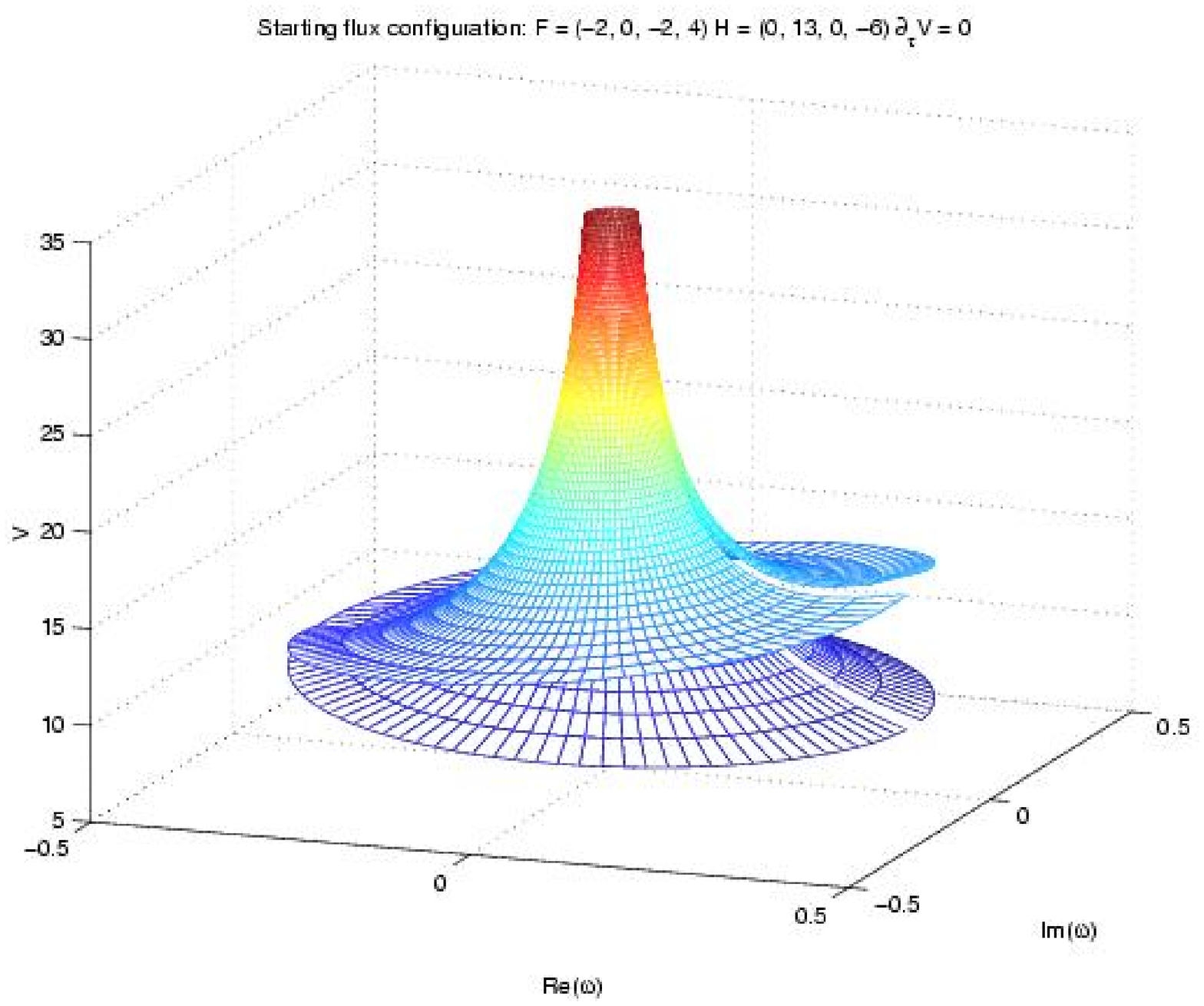}}
\caption{{\small \textsl{At the bottom of the spiral staircase there might be
a minimum of the type described in \cite{Giddings:2001yu}, or there might not be a
minimum. In these plots we use the convenient variable $\omega$, that relates
to $z$ as $\omega= (1-\ln|z^{-1/5}-1|/2)^{-1}e^{i \arg(z^{-1/5}-1)}.$}}}%
\label{fig:stair_bot}%
\end{figure}

Let us now study the staircase around $z=1$ in more detail. What might happen
is that if we go downwards we will eventually reach a minimum where the
contribution from $n\Pi_{1}$ (vanishing like $\xi$ at the conifold) and
$\Pi_{N}$ (vanishing like $\xi\ln\xi$) balance. This is nothing but the
minimum discussed in \cite{Giddings:2001yu} leading to a naturally large hierarchy. An
example is shown in upper plot of figure \ref{fig:stair_bot}. We need not, however, reach a
minimum at the bottom of the stairs. In some cases we would simply slide off
the staircase outwards from the conifold point, as the second picture in
figure \ref{fig:stair_bot} shows.

\bigskip

\subsection{Large complex structure}

\bigskip

Many of the features that we have found around the conifold point have
correspondences around the large complex structure (LCS) point, $z=0$. The large
complex structure is not as universal as the conifold behaviour among
different Calabi--Yaus, so we must be cautious when drawing conclusions from
the mirror quintic example. However, we expect the qualitative features we
list here to hold also in a more general setting.

\begin{figure}[tb]
\centering
\includegraphics[height=8cm]{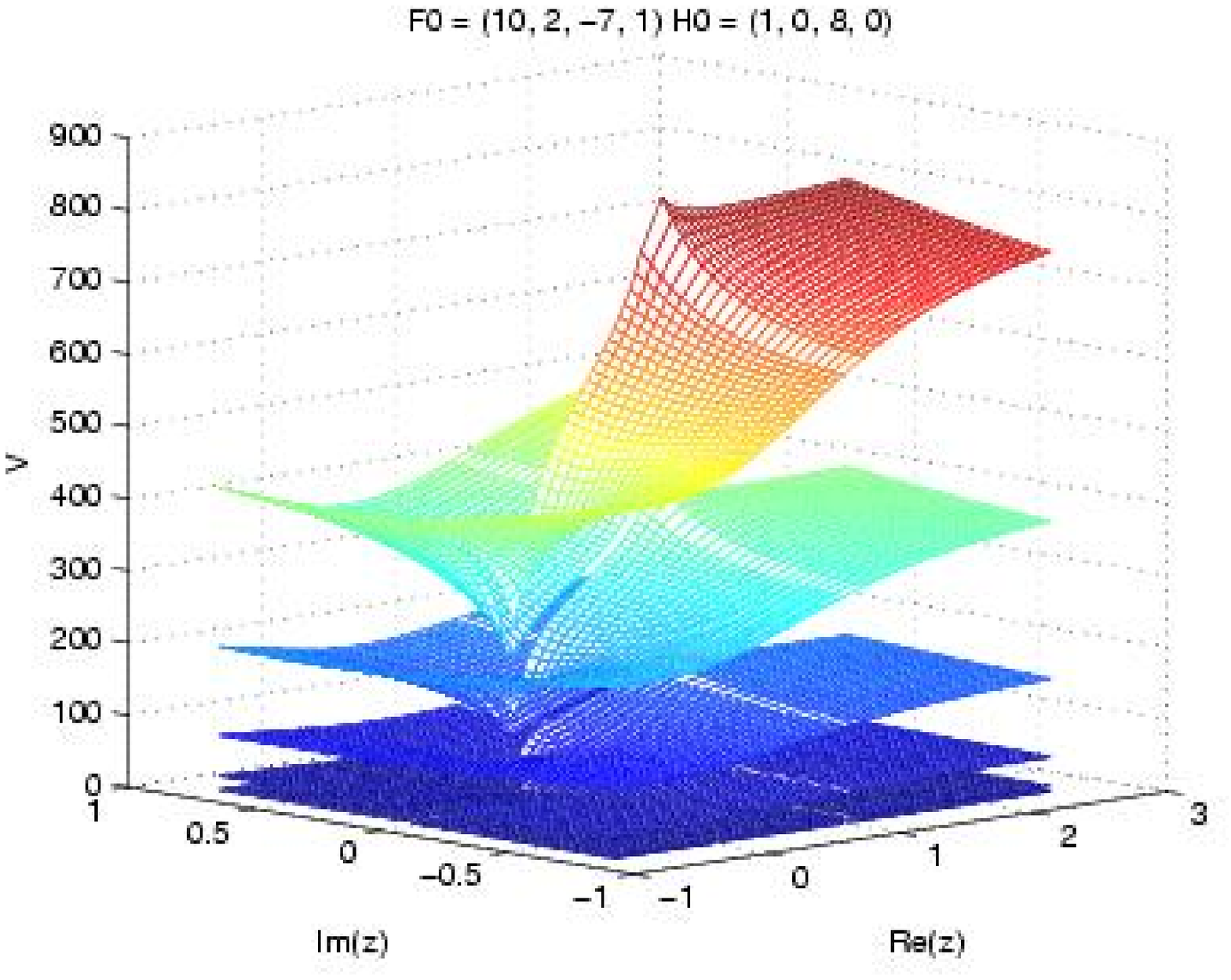} \caption{{\small \textsl{Spiral
staircase around the large complex structure point. The starting flux
configuration is $F_{0}$, $H_{0}$. We then change sheets with $T[0]$.}}}%
\label{fig:sks_spiral}%
\end{figure}

As for the conifold, there is a cut extending from the LCS point, giving us
another possibility of connecting sheets of the potential. Hence, there is a
spiral stairway encircling the LCS point, as shown in figure \ref{fig:sks_spiral}. 
Note that, since the monodromy
transformations $T[1]$ and $T[0]$ do not commute, the two staircases take us
to different levels of the potential.

Near the point of large complex structure we have, for large $t=\frac{5i}%
{2\pi}\ln(5z^{-1/5})$ \cite{Denef:2001xn},%
\begin{align}
\Pi_{1} &  \sim-\frac{5}{6}t^{3}\\
\Pi_{2} &  \sim-\frac{5}{2}t^{2}\\
\Pi_{3} &  \sim t\\
\Pi_{4} &  \sim1
\end{align}
and%
\begin{equation}
e^{-K}\sim\frac{20}{3}\left(  \operatorname{Im}t\right)  ^{3}%
\end{equation}
A straightforward calculation shows that, for general fluxes, $V\sim t^{5}$
and there is no minimum for large $t$:s. However, for the particular case when
$F_{1}=F_{2}=H_{1}=H_{2}=0$ we find, to leading order,%
\begin{equation}
\tau\sim-\frac{F_{3}}{H_{3}}+\frac{F_{4}H_{3}-F_{3}H_{4}}{H_{3}^{2}\bar
{\tilde{t}}}%
\end{equation}
where $\tilde{t}=\operatorname{Re}t+\frac{i}{\sqrt{3}}\operatorname{Im}t$. The
potential becomes%
\begin{equation}
V \sim \frac{ \left\vert \tilde{t} \right\vert ^{2} } { (\operatorname{Im}t)^4 }
 \frac{  \left( F_{4}H_{3}-F_{3}H_{4} \right)^2 }{ H_3^2 }
\left( 1 + 4\frac{ (\operatorname{Re}t)^2}{\left\vert \tilde{t} \right\vert ^{2}} + \mathcal{O}(\frac{1}{t}) \right).
\end{equation}
We see that the potential generically vanishes at the point of large complex
structure. As soon as we allow for any other non-zero fluxes, the zero will be
replaced by a spike and a spiral staircase, just as in the conifold case.

\bigskip

\section{Can we find infinite series?}

\bigskip

Our discussion above has not settled the question of whether there are
infinite series of continuously connected minima. Although we have not found
any infinite series in the particular example of the mirror quintic, we have
not given any argument to whether this might hold for a general Calabi--Yau. If
such series exist, and prevail in the full string theory, we would see large
effects on the topography of the landscape. Therefore we need to investigate
the question further.

There are arguments \cite{Ashok:2003gk,Denef:2004ze} that infinite series of
supersymmetric vacua always correspond to decompactification limits of the
effective four-dimensional theory. Naturally, such decompactified theories
would not be part of a landscape of four-dimensional theories.
These arguments are based on the fact that the tadpole condition
(\ref{eq:tadpole}) is positive definite when we lift the Type IIB
compactification to its F-theory correspondence. This only holds for
supersymmetric vacua. Here we also study non-supersymmetric vacua, so we need
a more general analysis.

To obtain an infinite series with fluxes of the form
\begin{align}
& F_{n}=F_{0}+nF_{L}\nonumber\\
& H_{n}=H_{L}%
\end{align}
we need fluxes $F_{L}$ and $H_{L}$ that have a minimum and that fulfill the condition%
\begin{equation}
\int_X F_{L}\wedge H_{L}=0.\label{eq:zero_wedge}%
\end{equation}
If this was not the case, an infinite series would eventually
violate the tadpole condition. As discussed in previous sections we have not
been able to find any infinite series of this form making use of monodromies
around the conifold point. We have also tried a more generalized approach to
the search for continuously connected infinite series, where we apply an
arbitrary monodromy $T$ to get to a sheet with a minimum, that is, we have an
effective set of fluxes given by%
\begin{align}
F &  =\left(  F_{N},0,0,...,0\right)  T\\
H &  =\left(  H_{1},H_{2},...,H_{N+1},0\right)  T.
\end{align}
In the case of the mirror quintic, $T$ would correspond to a combination of
$T[1]$:s and $T[0]$:s, but for a general Calabi--Yau we might have additional
monodromy matrices to choose from. While the first series of sheets have
the form%

\begin{align}
F &  =nF_{N}\left(  1,0,0,...,0\right)  +\mathcal{O}\left(  n^{0}\right)  \\
H &  =\mathcal{O}\left(  n^{0}\right)  ,
\end{align}
the new one is of the form%
\begin{align}
F &  =nF_{N}\left(  1,0,0,...,0\right)  T+\mathcal{O}\left(  n^{0}\right)  \\
H &  =\mathcal{O}\left(  n^{0}\right)  ,
\end{align}
where we must make sure that $T$ is such that $\mathcal{O}\left(
n^{0}\right)  T\sim\mathcal{O}\left(  n^{0}\right)  $. The picture to have in
mind is that we have at least two stairways, e.g. the one around the conifold
and the one around the large complex structure. At each floor of the conifold
stairway we go off in the direction $T$ to find a minimum. We then use
$T^{-1}$ to get back to the stairway to continue upwards to the next floor. 

\begin{figure}[t]
\centering
\includegraphics[height=12cm]{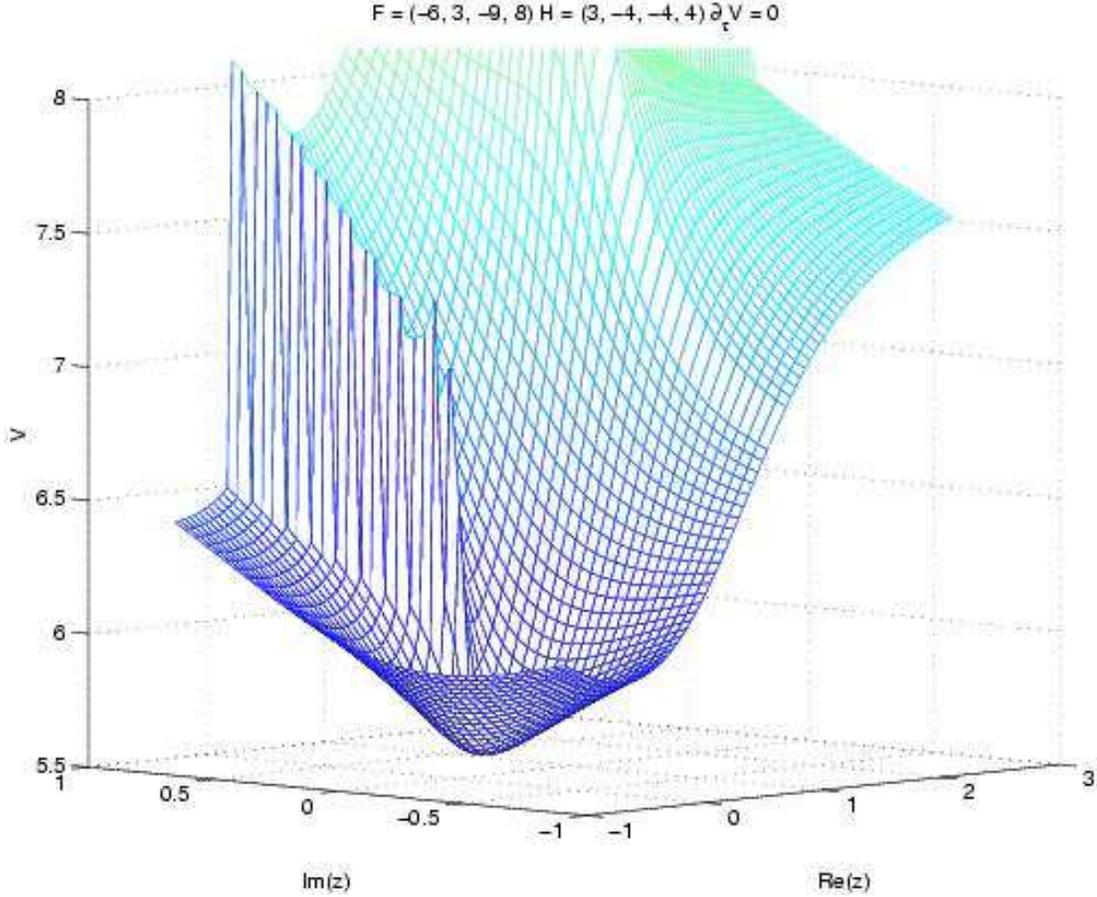} \caption{{\small \textsl{The scalar
potential $V$ for a flux configuration for which $\int_{X} F_{(3)}\wedge
H_{(3)} = 0$. From these fluxes it is possible to construct an infinite series
of minima. The minima are, however, not generally related by monodromies.}}}%
\label{fig:zero_wedge}%
\end{figure}

Unfortunately, we have not been able to find series of continuously connected
minima even in this generalized framework. However, if we relax the
requirement that the minima in the series need to be connected through
monodromies the situation changes. It is, in fact, very easy to find $F_{L}$
and $H_{L}$ satisfying (\ref{eq:zero_wedge}) and whose potential has a
minimum. We list a few examples in table 1. One of these
minima is also shown in figure \ref{fig:zero_wedge}.

\begin{table}[tbh]
\centering
\begin{tabular}
[c]{c|c}\hline
F & H\\\hline
$(-6,3,-9,8)$ & $(3,-4,-4, 4)$\\
$(4,2 ,8, 3)$ & $(8, 10, 12, -8)$\\
$(-10, 10, 1, 31)$ & $(2, -8, -7, 8)$\\
$(-5, 7, -5, -25)$ & $(2, 3, 5, 0)$\\
$(-18, 6, 20, 69)$ & $(2, 3, 14, -9)$\\
$(9, 4, -1, -26)$ & $(1, -17, 4, -3)$\\
$(-19, 11, -19, 18)$ & $(1, -5, -12, 11)$\\
$(0, -1, -13, -2)$ & $(5, 2, 16, -17)$\\\hline
\end{tabular}
\caption{{\small \textsl{A few flux configurations having $\int_{X}
F_{(3)}\wedge H_{(3)} = 0$ and whose corresponding potentials have a minimum.
This allows for infinite series of minima, albeit not (necessarily) connected
by monodromies. The potential corresponding to the first fluxes is plotted in
figure \ref{fig:zero_wedge}.}}}%
\label{tab:zero_wedge} 
\end{table}

To obtain the (possibly disconnected) series associated with this limiting
flux configuration, we need a symplectic transformation $S$ that takes us from
one minimum to the next in the series. For the first example in table 1 such a
transformation is given by%
\begin{equation}
S=\left(
\begin{array}
[c]{cccc}%
49 & -24 & 72 & -64\\
54 & -26 & 81 & -72\\
18 & -9 & 28 & -24\\
36 & -18 & 54 & -47
\end{array}
\right)  ,
\end{equation}
which has the required property of leaving the limiting fluxes $F_{L}%
=(-6,3,-9,8)$ and $H_{L}=(3,-4,-4,4)$ invariant, i.e. $F_{L}\cdot S=F_{L}$ and
$H_{L}\cdot S=H_{L}$. A series of minima where we continuously can go between
the minima would have a transformation $S$ that belong to the subgroup of the
symplectic group generated by the monodromies, i.e. the monodromy group.
Otherwise we have a series of minima where each minima sits on a different,
disconnected, piece of the landscape. Pictorially, we might think of these
disconnected pieces as different islands in the landscape. Minima on the same
island are connected by continuous paths, while we need to make discontinuous
jumps to move between the islands.

To be precise, it is enough that $S^{k}$, for some integer $k,$ is part of the
monodromy group in order for us to be able to find an infinite series of
continuously connected minima on the same island.
When we act with $S$ our minima might jump from island to island but
eventually our $S^{k}$ is such that it can be generated by acting with the
monodromy group on a previous minimum in the series. That is, our new minimum
is continuosly connected with a previous one.

Have we any guarantee that this always is the case? The question can be
phrased in terms of the index of the monodromy group as a subgroup of the
symplectic group. The index of a subgroup is the number of elements in the
group needed if action by the subgroup on these elements is supposed to
generate the full group. In other words it is the number of left cosets
corresponding to the subgroup in the full group. If the index is finite we can
be sure that there exists a finite $k$ as required above.

Unfortunately, the finiteness of the index of the monodromy group is an open
question, as discussed in \cite{Chen:2006}. There are however reasons to
believe that the index is infinite \cite{Zudilin}. Experimentally it seems as
if the matrices generated by the monodromy subgroup constitute a measure zero
set among all the symplectic matrices, even though there does not exist a
rigorous proof of this statement. This could indicate that the number of
elements needed to generate the full symplectic group would be, more or less,
the number of elements in the symplectic group, which of course is infinite.

If the index really is infinite, we can not be sure that the infinite series
we have generated will correspond to continuously connected series on the same
island. There could still be infinite series of minima where we stay on a
particular island, but which of the series corresponding to the examples
listed above that is of this form has to be checked case by case.
Unfortunately it is not obvious how to do this in a systematic and efficient
way, and we find it intriguing that the topography of the landscape is
sensitive to these unresolved mathematical problems.

The possibility of different vacua at different regions of space, leads to the
existence of domain walls. A domain wall in four dimensional space-time can be
thought of as a five brane wrapped around some combination of three cycles on
the internal manifold. The different vacua on the two sides of the
domain wall have fluxes that differ in a way given by the way the brane is
wrapped \cite{Gukov:1999ya}. It can also be shown that the effective four dimensional tension of
the domain wall is bounded from below by the absolute value of the change in
the superpotential when we go from one side of the domain wall to the other
\cite{Ceresole:2006iq}. In this way, we can in principle construct domain walls separating
any two different flux vacua, and get an estimate of their tension.

For a given Calabi--Yau, such as the mirror quintic, our results seemingly
imply that there are actually two types of domain walls. The first kind have
fluxes relating two vacua connected through an element belonging to the
monodromy group, and the second kind have fluxes that can not be related in
this way. For the first type we can derive a profile depending on the complex structure moduli
interpolating between the two regions. For the second type of domain wall our
theory does not allow us to do this. Another way to put this is to say that
the first type of domain walls separate minima situated on the same island,
while the second type separate minima on different islands.

In the full string theory with all moduli at our disposal, we would expect to
be able to derive profiles of all possible domain walls. In other words,
generalizing our procedure to the full string theory, we can construct bridges
connecting the islands of the landscape. Thus, all the different islands is
actually part of the same connected landscape.

It is possible that when a more complete understanding of string theory is
reached, one will find the separation into two types of domain walls
artificial, and thus that all islands are really only parts of the same continent.

\bigskip

\section{Conclusions and outlook}

\bigskip

In this paper we have explored the string landscape for flux compactified type
IIB string theory. We have in particular studied the occurrence of multiple
vacua, and found that there are good reasons to expect that series of closely
positioned vacua are rather common. As an example, we presented a series of 17
consecutive minima related trough conifold monodromies on the mirror quintic.
We have furthermore demonstrated that periodic series of minima, differing
only by the value of the axion, are a general feature of these models.

We have also argued that there are interesting features of the string
landscape, related to the distribution of minima, which depend crucially on
the mathematical properties of the monodromy subgroups of the symplectic
group. Through an explicit example using the mirror quintic, we have showed that
infinite series of minima exist, but we have not found any such series where
the minima are connected by monodromies. In other words, we have not found
infinite series where we continuously can move from one minimum to another. The
question of if and when this is possible is intimately connected with the
mathematically unresolved problem of the finiteness of the index of the
mondoromy subgroup in Sp($N$,$\mathbb{Z}$). 

In our work we have focused our attention to the complex structure moduli, but
it is important to investigate what happens to our series when we embed them
into more realistic models where also the K\"{a}hler moduli are stabilized. In
case of the popular KKLT scenario \cite{Kachru:2003aw} we assume a superpotential of the form%
\begin{equation}
W=W_{0}\left(  z\right)  +Ae^{i\rho},
\end{equation}
where the fixing of the complex structure moduli is assumed to be independent
of the fixing of the K\"{a}hler moduli. In the original KKLT proposal, it was
assumed that the lifting of the resulting AdS vacua to de Sitter, is achieved
through adding anti-D3 branes that also break the supersymmetry. An
alternative discussed in \cite{Saltman:2004sn} is to instead consider non-zero
F-terms, in line with what we have allowed in the no-scale case. It is easy to
see that the stabilization in the K\"{a}hler direction requires that
$-A<W_{0}<0$, which will have important consequences for our series of minima.
We have found that the superpotential $e^{K/2}W$ scales with $\sqrt{n}$, and
that the potential $V\rightarrow\infty$ as $n$ grows. Since $W_{0}$ grows
along the series, \ we will eventually find ourselves outside of the interval,
and the theory destabilizes in the K\"{a}hler direction. Even before this
happens, the large value of $W_{0}$ will bring us into a regime where the
perturbative corrections to the K\"{a}hler potential will become important, as discussed in
\cite{Conlon:2005ki}. It is important to investigate the fate of our series of
minima in this more general setting.

An obvious extension of our work would be the study of the detailed form of
the series of minima -- infinite or not -- in view of applications to the
early universe and inflation. What are the typical potential barriers and
domain wall tensions in a series of minima? Is resonant tunneling a naturally
occurring phenomena or is fine tuning needed?  Could some of our series serve
as a basis for a model of chain inflation? We hope to return to these and
other problems in future publications.

\bigskip

\section*{Acknowledgments}

The work was supported by the Swedish Research Council (VR). We thank Wadim
Zudilin, Torsten Ekedahl and Ernst Dieterich for useful discussions.

\end{document}